\documentclass{aa501}
\usepackage{graphicx}
\usepackage{natbib}

\begin{document}

   \title{Colour Evolution of Disk Galaxy Models from z=4 to z=0}

   \author{P. Westera \inst{1} \and M. Samland\inst{1} \and
           R. Buser\inst{1} \and O. E. Gerhard\inst{1} }

   \offprints{M. Samland}

   \institute{Astronomisches Institut der Universit\"at Basel,
             Venusstrasse 7, CH-4102 Binningen, Switzerland\\
             \email{samland@astro.unibas.ch, westera@astro.unibas.ch}
             }

   \date{Received <date>; accepted <date>}

   \abstract{ We calculate synthetic $UBVRIJHKLM$ images, integrated
   spectra and colours for the disk galaxy formation models of
   \citet{samland_02}, from redshift $z=4$ to $z=0$. Two models are
   considered, an accretion model based on $\Lambda$CDM structure
   formation simulations, and a classical collapse model in a dark
   matter halo. Both models provide the star formation history and
   dynamics of the baryonic component within a three-dimensional
   chemo-dynamical description. To convert to spectra and colours, we
   use the latest, metallicity-calibrated spectral library of
   \citet{westera_02}, including internal absorption. As a first
   application, we compare the derived colours with Hubble Deep Field
   North bulge colours and find good agreement. With our model, we
   disentangle metallicity effects and absorption effects on the
   integrated colours, and find that absorption effects are dominant
   for $z<1.5$. Furthermore, we confirm the quality of $m_{K}$ as a
   mass tracer, and find indications for a correlation between
   $(J-K)_{0}$ and metallicity gradients.
    }

   \maketitle

\section{Introduction}

   Today, it is possible to observe galaxies out to high redshift and
   to study how they form and evolve. Long exposures in different
   wavelength bands result in images with very faint limiting
   magnitudes, such as the Hubble Deep Field
   \citep[HDF][]{williams_96} and its NICMOS counterpart
   \citep{thompson_99}, or the FORS deep field \citep{appenzeller_00},
   to name but a few. In these deep fields, we can see galaxies back
   to epochs shortly after their formation. Together with ground-based
   observations, these data provide morphological and photometric
   information on the evolution of disk galaxies as a function of
   redshift \citep{vogt_96, roche_98, lilly_98, simard_99}. At
   redshifts $z>2$, the deep fields reveal a wide range of galactic
   morphologies with considerable substructure and clumpiness
   \citep{pentericci_01}. From redshift $z=2$ to $z=1$, massive
   galaxies seem to assemble \citep{kajisawa_01}, while for $z<1$ most
   of the Hubble type galaxies show only little or no evolution
   \citep{lilly_98}. These results are obtained from still small
   samples of galaxies, but with the new large telescopes much more
   information about high redshift galaxies will be available in the
   future.

   However, for understanding of the galaxy formation process also
   theoretical models are needed. Modern galaxy formation models,
   based on the hierarchical structure formation scenario, predict
   halo formation histories and the assembly of the baryonic matter
   inside these halos \citep{nagamine_01, pearce_01, cole_00,
   navarro_00, hultman_99}, but the spatial resolution of these
   simulations is not sufficient to describe the formation of galaxies
   in detail. This can be done either in the framework of
   semi-analytical models \citep{cole_00, firmani_00, diaferio_99,
   mo_98}, with hybrid models \citep{boissier_01, jimenez_98} or with
   dynamical models that simulate the formation and evolution of
   single galaxies \citep{bekki_01, williams_01, berczik_99,
   samland_97, steinmetz_95a, katz_91}. In order to compare the models
   with the colours and magnitudes of real galaxies, realistic
   transformations of the models into spectral properties are
   needed. Only through transformation into spectra and colours, can
   the galactic models be compared with observations and thereby be
   confirmed or refined.

   Recent applications of such transformations include
   \citet{gronwall_95}, who use the \citet{bruzual_93} Galaxy
   Isochrone Spectral Synthesis Evolution Library (GISSEL93) code to
   derive integrated spectra for their models of galaxies of different
   spectral type, and then obtain final spectral energy distributions
   by adding a simple reddening with a constant $E_{B-V}$ of 0.1. The
   GISSEL93 code was also used by \citet{roche_96} for their
   non-evolving and pure luminosity evolution models of different
   galaxy types, but they use an absorption coefficient that is
   proportional to the star formation rate (SFR) divided by the galaxy
   mass. \citet{campos_97} also use the GISSEL93 code and a constant
   absorption for their spiral and early type luminosity evolution
   models. To take account of the cosmology, they used the K
   corrections of \citet{metcalfe_91}. The 1999 version of GISSEL,
   combined with the BaSeL 2.2 \citep{lejeune_97, lejeune_98}
   semi-empirical stellar spectral energy distribution (SED) library
   was used by \citet{kauffmann_98a, kauffmann_98b} for their
   semi-analytical models. \citet{jimenez_98, jimenez_99} use their
   own isochrones and Kurucz 1992 \citep{buser_92} SEDs, complemented
   with atmosphere models of their own, in their low surface
   brightness disk galaxy models. \citet{contardo_98} interpolate the
   colour evolution of the underlying stars from a grid of theoretical
   colour evolutionary tracks and then apply a K correction. In all
   these models no correction for internal dust absorption is made.

   In this paper, we combine the disk galaxy formation models of
   \citet{samland_02} with the latest metallicity-calibrated stellar
   SED library \citep{westera_02} and galaxy evolutionary code
   \citep{bruzual_00}, including the spatially resolved internal
   absorption obtained from the three-dimensional distribution of gas
   in these models. We obtain $UBVRIJHKLM$ images and spectra
   (intrinsic and redshifted) of the model galaxies. The redshifted
   spectra include the Lyman line blanketing and Lyman continuum
   absorption by absorption systems at cosmological distances using
   the formulae given by \citet{madau_95}. Comparison of the model
   galaxies with bulge observations in the Hubble Deep Field North
   (HDF-N) shows good agreement, confirming our approach. We can
   disentangle different effects on the spectral properties of a model
   galaxy, such as from metallicity and internal absorption, by
   artificially blinding these contributions out, and then
   recalculating the spectral properties.

   The outline of the paper is as follows: In Section~\ref{chapter2},
   the galaxy models are briefly described \citep[for a detailed
   description, see][]{samland_02}. In Section~\ref{chapter3}, we
   discuss the programme used for the transformation into colours and
   spectra, and in Section~\ref{chapter4}, we present our first
   results and a comparisons with empirical (HDF-N) data. In the last
   section, conclusions are drawn, and an outlook on further work is
   given.

\section{Short description of the new galaxy evolution models}
\label{chapter2}

   The observations of high redshift galaxies of interest here provide
   magnitudes, colours and some information about morphology
   (asymmetry and concentration parameter). Interpreting these data
   fully requires detailed models for galactic evolution. In this
   paper, we want to show, that a dynamical multi-phase galaxy model
   provides the necessary physical information to interpret the high
   redshift data. For this purpose, we use the 3-dimensional
   chemo-dynamical models described in detail in the companion paper
   \citep{samland_02}. Here, we only summarize briefly their main
   properties. These models include cosmological initial conditions,
   dark matter, stars and the different phases of the interstellar
   medium (ISM), as well as the feedback processes which connect the
   ISM and the stars.

   We use two different models, both describing the formation and
   evolution of a disk galaxy in a $\Lambda$CDM universe
   ($H_{0}=70$~km/s/Mpc, $\Omega_{0}=0.3$, $\Omega_{\Lambda}=0.7$, and
   $M_{baryon}/M_{dark}=1/5$). The spin parameters of the model
   galaxies are chosen to be $\lambda=0.05$ \citep{gardner_01,
   vandenbosch_98, cole_96, steinmetz_95b, barnes_87} and we follow
   the evolution of both models from $z=4.85$ (corresponding in this
   cosmology to a universal age of $1.2$~Gyr) until $z=0$
   ($13.5$~Gyr).

   The first model, which we call the collapse model, is an extreme
   case which starts with an extended halo of $250$~kpc radius and
   which has a total mass of $1.8 \cdot 10^{12}
   \mathrm{M}_{\sun}$. Initially the baryonic and dark matter is
   distributed according to the density profile proposed by
   \cite{navarro_95}. We assume that only the baryonic matter can
   collapse, similar to an Eggen, Lynden-Bell, and Sandage scenario
   \citep{eggen_62}. We use this model mainly as a reference to
   highlight the differences to a second more realistic,
   cosmologically motivated model.

   This second model, from now on called the accretion model, is
   characterized by a slowly growing dark halo with a continuous gas
   and dark matter infall. The time dependent accretion rate is
   derived by averaging 96 halo merging histories from cosmological
   N-body simulations from the VIRGO-GIF project
   \citep{kauffmann_99}. In this scenario, the dark halo grows slowly
   from a radius of $15$~kpc at $z=4.85$ to $250$~kpc at $z=0$. We
   assume that, at $z=4.85$, the baryonic matter outside the $r_{200}$
   radius consists of ionized primordial gas. The accreted gas can
   cool, forms clouds, dissipates kinetic energy and finally collapses
   inside the dark halo. The collapse is delayed by the feedback
   processes and a galaxy with an extended disk forms. This is in
   agreement with the result of \citet{weil_98}, that the formation of
   disc galaxies requires feedback processes which prevent gas from
   collapsing until late epochs.

\begin{figure}
     \includegraphics[width=\columnwidth]{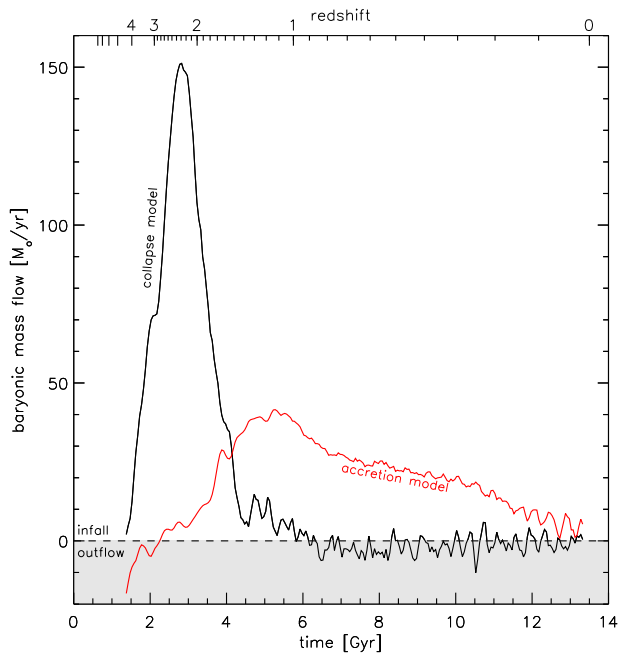}
     \caption{Baryonic mass flow into resp. out of a sphere of
     $20$~kpc radius surrounding the galaxy centres of the accretion
     and the collapse models. Negative mass flows (grey shaded region)
     correspond to net outflows.}
     \label{bild01} 
   \end{figure}

\begin{figure*}
     \includegraphics[width=\textwidth]{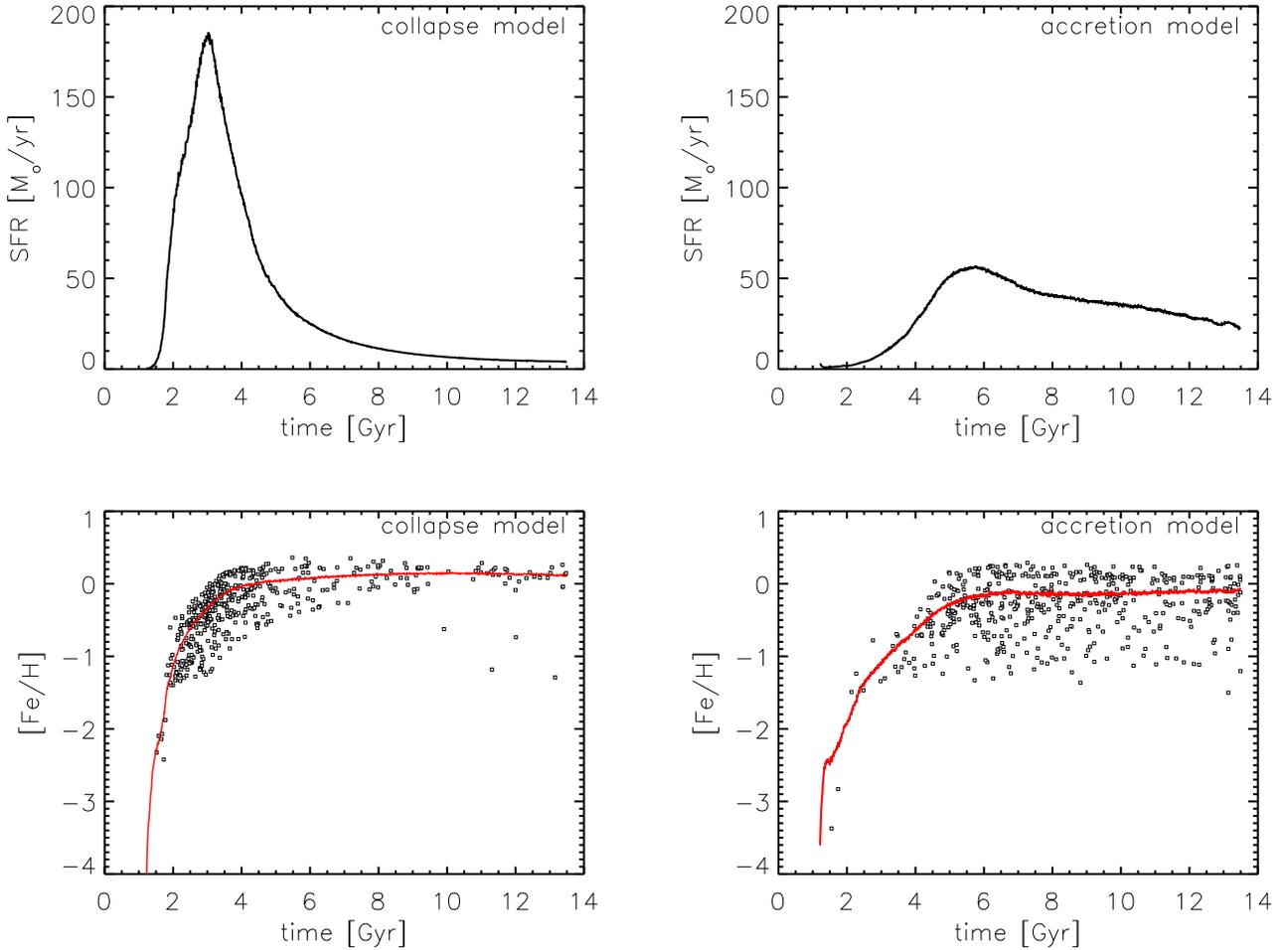}
     \caption{Upper panels: the star formation rates of the collapse
     model (left), and the accretion model (right). Lower panels: the
     age-metallicity distributions and average metallicities of the
     stellar particles. The small squares show a representative sample
     of stellar particles.}
     \label{bild02}
   \end{figure*}

   In the collapse model, the infall of baryonic matter into the
   innermost $20$~kpc of the dark halo is determined only by
   dissipation and feedback processes between stars and ISM. The black
   line in Fig.~\ref{bild01} represents the baryonic mass flow into
   resp. out of a sphere of $20$~kpc radius surrounding the model
   galaxy centre. The collapse model shows an early mass infall that
   ends more or less at $z=1$. Later there is some in and outflow, but
   this does not change the mass of the galaxy significantly. As this
   $20$~kpc region is responsible for most of the star formation (SF),
   the total SFR (Fig.~\ref{bild02}, upper left panel) is strongly
   correlated with the collapse time, and thus peaks very early at $z
   \simeq 2$ (corresponding to an age of $\sim 3$~Gyr). The modest SF
   from $z=1$ until the present epoch, is maintained by the gas return
   from long lived main sequence stars entering the planetary nebula
   phase. For the colour evolution of a galaxy it is important to know
   the SF and the enrichment history.  The lower left panel of
   Fig.~\ref{bild02} shows the metallicity distribution and the
   average metallicity of the stellar particles as a function of the
   time when they were born. In the first $\sim 1.5$~Gyr of the
   simulation, the metallicity shoots up from $\sim -4$ dex to around
   solar. From this point on, it stays constant, reaching not much
   more than $\sim 0.1$~dex at the present epoch. This can be
   explained by the fact that after the first $\sim 1.5$~Gyr, the bulk
   of the SF, and hence of the gas enrichment, is
   completed. Morphologically, the outcome of the collapse model is an
   early-type disk galaxy.

   In the accretion model, the galaxy forms in a smoother way. The
   mass flow into the inner $20$~kpc is shown in Fig.~\ref{bild01} by
   the grey (in the colour version: red) line. It has a maximum at
   $z=1.1$, but remains significant until the present epoch, because
   of the steady accretion of baryonic (and dark) matter. Therefore,
   we expect a larger mixture of stellar populations of many different
   ages and metallicities, compared to the collapse model. The SFR in
   the accretion model (Fig.~\ref{bild02}, upper right panel) peaks at
   around $z=1$ ($\sim 5.75$~Gyr after the big bang), and stays high
   until $z=0$. In analogy to the collapse model, the average stellar
   particle metallicity ${\rm [Fe/H]}$ of this model (shown in the
   lower right panel of Figure~\ref{bild02}) increases most steeply
   during the phase of maximum SF. It also starts at ${\rm [Fe/H]}
   \simeq -4$, and reaches its present value of $\sim -0.1$~dex at $z
   \simeq 1$. The accretion model galaxy forms from inside-out and
   from top-to-bottom, with the halo as the oldest component, followed
   by the bulge and the disk. At $z=1$, the galaxy begins to form a
   bar which later turns into a triaxial bulge. This model nicely
   produces a barred disk galaxy, and since it uses more realistic
   cosmological initial conditions, we shall in the following
   concentrate on this model.

\begin{figure*}
     \includegraphics[width=\textwidth]{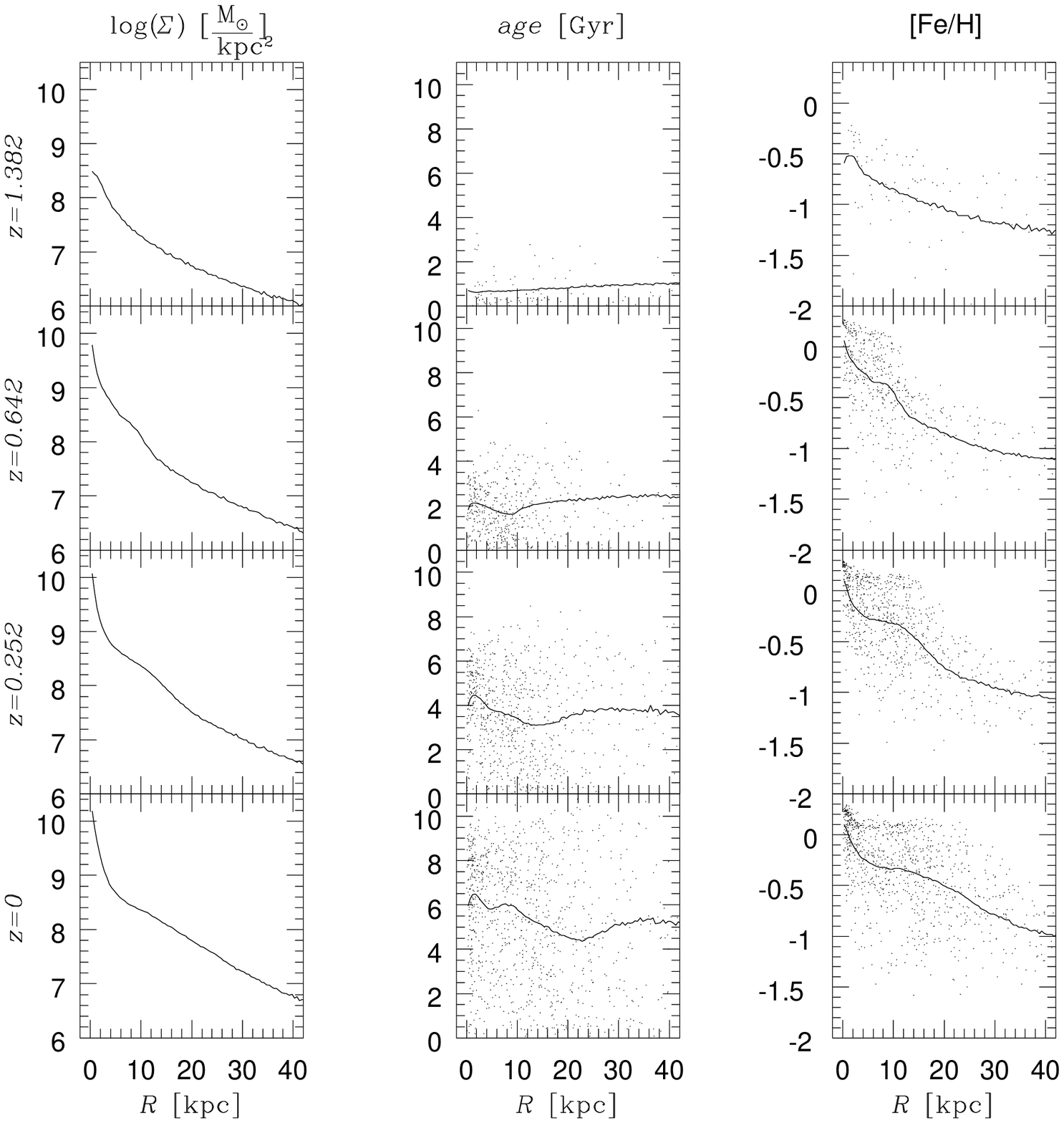}
     \caption{Surface density profiles (left), stellar particle age
     (middle) and ${\rm [Fe/H]}$ distribution (right) of the accretion
     model at four different redshifts (from top to bottom: $z=1.382$,
     $0.642$, $0.252$ and $0.000$, corresponding to universe ages of
     $4.5$, $7.5$, $10.5$ and $13.5$~Gyr). The lines represent the
     mean profiles, and the dots show a representative sample of
     stellar particles.}
     \label{bild03}
   \end{figure*}

   Fig.~\ref{bild03} shows the radial profile of the (stellar) mass
   surface density, the ages of the stellar particles, and the stellar
   metallicity, at four redshifts ($z=1.382$, $0.642$, $0.252$ and
   $0$, corresponding to universe ages of $4.5$, $7.5$, $10.5$ and
   $13.5$~Gyr). The profiles (the lines in Fig.~\ref{bild03}) were
   calculated by first projecting the respective quantities on the
   disk plane, then determining the point of the highest (stellar)
   mass concentration in this projection, and in the end averaging
   (mass-weighted) the projections over rings surrounding this
   point. To show the spread in stellar particle age and metallicity,
   a representative sample of stellar particles is shown as dots.

   In the surface density profile, one can see a clear bulge and a
   bump appearing at around $6$~Gyr at a radius of $\sim 10$~kpc,
   indicative of the bar \citep{lerner_99, efstathiou_82}. These
   features are clearly visible in the images in
   Section~\ref{chapter4}. In stellar age, not much of a radial
   dependence can be seen, apart maybe from a small negative gradient
   in the inner region, whose outer limit (thus the minimum) slowly
   wanders outwards from $8$~kpc at $z=0.642$ to $20$~kpc today, due
   to an inside-out forming disk. In metallicity, there is an evident
   negative gradient, that stays nearly constant from $z=0.6$ on. This
   is due to a combination of the metallicity gradient of the disk,
   which, due the redistribution from the bar, is flat out to $\sim
   10$~kpc (hence the bump there) and drops further outwards, and the
   shallow halo gradient. More quantitatively, the average metallicity
   reaches solar in the centre, with the most metal-rich stellar
   particles reaching ${\rm [Fe/H]=+0.3}$ while at $40$~kpc from the
   centre, the average metallicity has dropped to ${\rm [Fe/H] \simeq
   -1}$. The important results which we need in
   Subsection~\ref{chapter4.4} are:

   \begin{enumerate} 
     \item The surface density drops with radius at all times, and
     from $z=0.6$ on, a very steep gradient in the inner few kpc and a
     bump at around 10 kpc reflect the existence of a bulge and a bar.
     \item The stellar age distribution shows almost no gradient.
     \item The metallicity drops with increasing projected radius,
     also showing a bump at around $10$~kpc.
   \end{enumerate}

   The output quantities of interest (which are the input quantities
   for the programme which calculates the spectral properties) are the
   following: $614500$ stellar particles, each with its position in
   $x$, $y$, and $z$, initial mass, age, and metallicity, as well as
   the gas density on a three-dimensional grid covering the galaxy out
   to where the gas density is negligible ($100$~kpc). We use the gas
   density to calculate the internal dust absorption in the model
   (Section~\ref{chapter3}). All quantities are followed during the
   whole galactic evolution. With these data we determine the
   evolution of the brightness, colours, metallicities and the
   structure of the model galaxy. In comparison with the observations,
   we can use the results to interpret the high redshift data but also
   to improve the galactic models and to learn more about the
   processes that strongly influence the galactic evolution.

\section{From theoretical quantities to colours and spectra}
\label{chapter3}

   To derive 2-dimensional colour ($UBVRIJHKLM$) images from the
   distributions of stars and gas, we proceeded in the following way:

   First, a library of simple stellar population (SSP) spectra was
   produced. With the Bruzual and Charlot 2000 Galaxy Isochrone
   Spectral Synthesis Evolution Library (GISSEL) code
   \citep{charlot_91, bruzual_93, bruzual_00}, integrated spectra
   (ISEDs) of populations were calculated for a grid of population
   parameters consisting of $8$ metallicities (${\rm [Fe/H]} =
   -2.252$, $-1.65$, $-1.25$, $-0.65$, $-0.35$, $0.027$, $0.225$, and
   $0.748$) and 221 SSP ages ranging from $0$ to $20$~Gyr. As input,
   we used Padova 2000 isochrones \citep{girardi_00}. For the highest
   and the lowest metallicity, where no Padova 2000 isochrones are
   available, we used Padova 1995 isochrones \citep{fagotto_94,
   girardi_96}. The spectral library used was the BaSeL 3.1 ''Padova
   2000'' \citep{westera_02, westera_01} stellar library. This library
   is able to reproduce globular cluster colour-magnitude diagrams in
   combination with the Padova 2000 isochrones for all metallicities
   from ${\rm [Fe/H]} = -2.0$ to $0.5$, because it was calibrated (in
   a metallicity-dependent way) for this purpose, and does a similarly
   good performance on integrated SSP spectra. Colour differences with
   template empirical spectra from \citet{bica_96} amount to a few
   $100$th of a magnitude only \citep{westera_02, westera_01}. A
   Salpeter initial mass function (IMF) with cutoff masses of
   $0.1~\mathrm{M}_{\sun}$ and $50~\mathrm{M}_{\sun}$ was chosen in
   accordance with the galaxy models. The spectra of this ISED library
   contain fluxes at $1221$ wavelengths from $9.1$~nm to $160$~$\mu$m,
   comfortably covering the entire range where galaxy radiation from
   stars is significant.

   After choosing the viewing direction and the size (up to $160
   \times 160$ pixels) and resolution for the ''virtual CCD camera'',
   the stellar particles are grouped into pixels. For each stellar
   particle, the spectrum is (flux point by flux point) interpolated
   from the ISED library. For metallicities lower than the range
   covered by the library (some stellar particles have metallicities
   down to ${\rm [Fe/H]} = -4.0$, but none have metallicities above
   the library range), the spectra for the lowest metallicity (${\rm
   [Fe/H]} = -2.252$) were used. This should not pose any problems, as
   trends of spectral properties with metallicity are expected to
   become weak below ${\rm [Fe/H]} = -2.0$, and these
   lowest-metallicity stellar particles become negligible in number
   very soon ($\sim 0.5$~Gyr after the beginning of the simulation).

   The spectrum is reddened as follows. According to
   \citet{quillen_01} we assume a dust-to-gas ratio which is
   proportional to the metallicity of the gas. The absorption
   coefficient $A_{V}$ can be expressed as
   \begin{equation}
     A_{V} = \frac{\mathrm{pc}^{2}}{15~\mathrm{M}_{\sun}} \int_{\rm
     line~of~sight} \rho_{Z}(r) dr
   \end{equation} 
   with the metallicity-weighted gas density $\rho_{Z}(r) =
   \rho_{gas}(r) Z(r) / Z_{\sun}$. Only the cold cloud medium is
   assumed to contain dust. For each stellar particle, $\rho_{Z}(r)$
   is integrated along the line of sight to derive the absorption
   coefficient $A_{V}$. The spectrum of the stellar particle is then
   reddened using the extinction law of \citet{fluks_94}.

\begin{figure}
     \includegraphics[width=\columnwidth]{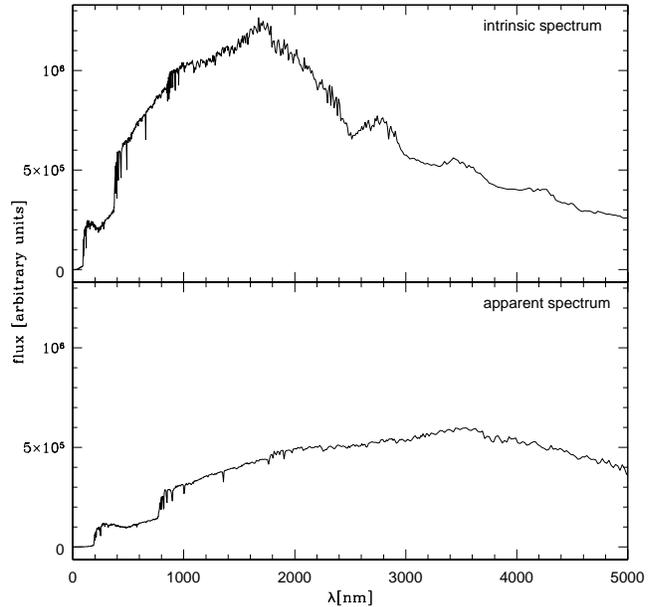}
     \caption{Intrinsic and apparent (redshifted) spectra of the
     accretion model at $z=1.065$ (corresponding to a universe age of
     $5.5$~Gyr).}
     \label{bild04}
   \end{figure}

   All the spectra of stellar particles from the same pixel are added
   up to give the integrated absolute spectrum of the pixel, which is
   then redshifted and dimmed according to the distance modulus $m-M$
   \citep{carroll_92}.
   \begin{eqnarray} 
     (m - M) = 25 + 5 \log \left( \frac{c (1+z)}{H_0} \right)
     \nonumber \\ + 5 \log \left( \int_{0}^{z} {dz' \over
     \sqrt{(1+z')^2(1+\Omega_Mz')- z'(2+z')\Omega_\Lambda}} \right).
   \end{eqnarray}
   In a next step, we correct the spectra for Lyman line blanketing
   and Lyman continuum absorption using the formulae given by
   \citet{madau_95} for QSO absorption systems. Finally, apparent
   $UBVRIJHKLM$ colours and magnitudes are calculated for each pixel
   through synthetic photometry. At the same time, the absolute (rest
   frame) spectra and the apparent spectra of all the pixels are added
   up to derive the absolute and apparent integrated spectra of the
   galaxy. An example of such an integrated spectrum is shown in
   Fig.~\ref{bild04}. It shows the intrinsic (upper panel) and the
   apparent (lower panel) spectra of the accretion model galaxy at
   $z=1.065$ from $0$~nm to $5000$~nm. These spectra differ from
   previous model spectra, because they take into account the
   three-dimensional metallicity and age distribution of the stars and
   the intrinsic dust absorption. However, as one can see from these
   spectra, gas emission lines are not implemented. Including HII
   regions, as well as planetary nebulae and supernovae, will be one
   of the next steps in improving the programme.

   On these integrated spectra, synthetic photometry is performed
   too. At the moment, the spectra of individual pixels or stellar
   particles are not stored. The final output quantities of the
   programme are:

   \begin{enumerate} 
     \item 2-dimensional $UBVRIJHKLM$ images of the model galaxy
     including the effect of internal absorption in apparent
     (redshifted and corrected for the distance modulus and Lyman line
     blanketing and by continuum absorption) magnitudes of up to $160
     \times 160$ pixels, as seen from a freely chosen angle.
     \item the integrated intrinsic spectrum of the entire galaxy plus
     integrated intrinsic colours and absolute magnitudes.
     \item the integrated apparent spectrum of the entire galaxy plus
     integrated redshifted colours and apparent magnitudes.
   \end{enumerate}

   We also included the possibility to account for Galactic foreground
   reddening, but this option only makes sense for specific
   applications, where the foreground reddening is known. In this
   work, it is not used.

   All these quantities were calculated for both the accretion and the
   collapse model, at universe ages from $1.5$~Gyr ($0.3$~Gyr after
   the beginning of the simulations) to $13.5$~Gyr (the present day)
   in steps of $0.5$~Gyr, and from three different directions:
   face-on, inclined by $60^{\circ}$, and edge-on. The size of a pixel
   was chosen to be $0.5$~kpc. At the moment higher resolution makes
   no sense, as the galaxy model has a resolution of only
   $0.37$~kpc. The entire ''camera'' was chosen $160 \times 160$
   pixels wide, thus representing a field of view $80 \times 80$~kpc
   wide.

   To study metallicity effects, the face-on and inclined images and
   spectra were calculated for the accretion model again, but
   assigning solar metallicity to each stellar particle. The
   differences between the regular accretion model and this model
   should therefore purely reflect metallicity effects, allowing us to
   estimate the error that is made in models using solar
   metallicity. For the sake of simplicity, we will from now on call
   this the solar metallicity model.

   Analogously, to identify absorption effects, the same photometric
   properties were calculated for the accretion model without internal
   absorption. Thus, the differences between the regular model and
   this one should reflect absorption effects, or the error that is
   made in models that do not include internal absorption. This model
   will be called the absorptionless model.

   Our synthetical photometric data have been produced in the
   Johnson-Cousins $UBVRIJHKLM$ system, but of course, they can in
   principle be produced in any system with known passband response
   functions. Other spectral features, such as line strength indices
   from the integrated pixel/galaxy spectra can also be derived.

\section{Results}
\label{chapter4}

\begin{figure*}
     \includegraphics[width=\textwidth]{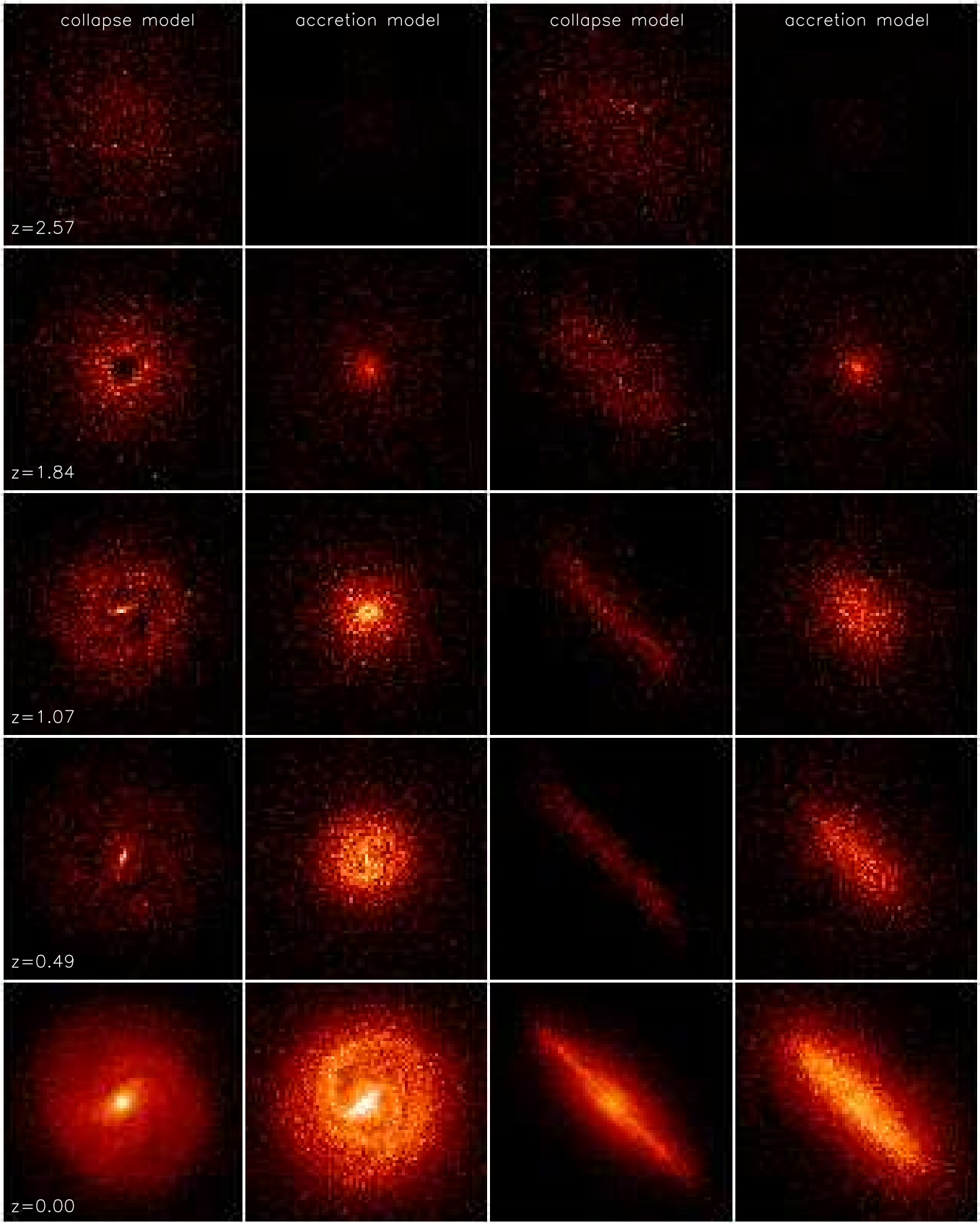}
     \caption{$U$ band evolution of the collapse and the accretion
     models. First column: collapse model, face-on, second column:
     accretion model, face-on, third column: collapse model, edge-on,
     fourth column: accretion model, edge-on. The wavelength ranges
     that are shifted into the $U$ band here correspond to ''bands''
     with effective wavelengths (from top to bottom) $103$~nm,
     $129$~nm, $177$~nm, $246$~nm, and $367$~nm ($U$ band).}
     \label{bild05}
   \end{figure*}

\begin{figure*}
     \includegraphics[width=\textwidth]{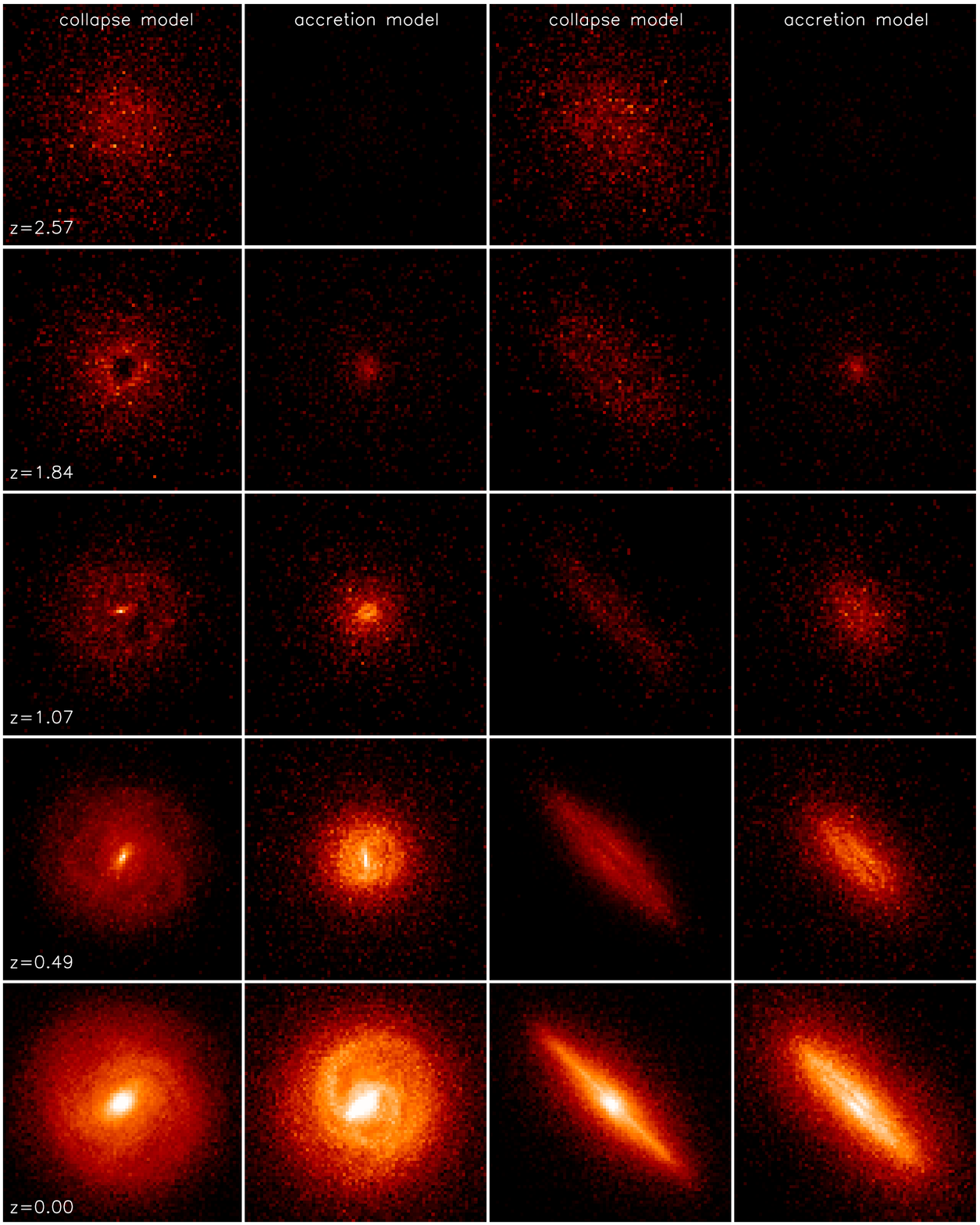}
     \caption{$V$ band evolution of the collapse and the accretion
     models. First column: collapse model, face-on, second column:
     accretion model, face-on, third column: collapse model, edge-on,
     fourth column: accretion model, edge-on. The wavelength ranges
     that are shifted into the $V$ band here correspond to ''bands''
     with effective wavelengths (from top to bottom) $153$~nm,
     $192$~nm, $263$~nm, $366$~nm ($\sim U$) and $545$~nm ($V$ band).}
     \label{bild06}
   \end{figure*}

\begin{figure*}
     \includegraphics[width=\textwidth]{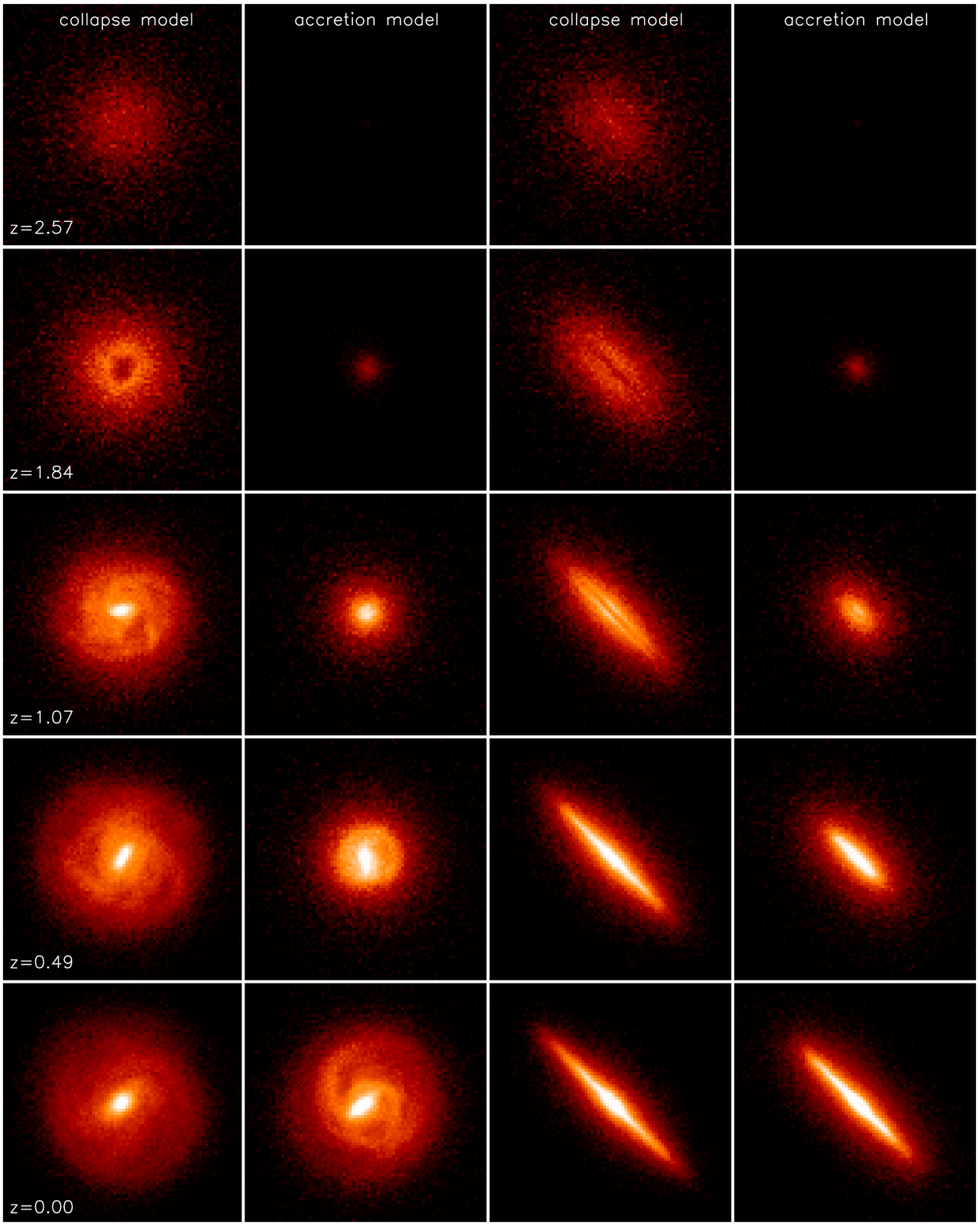}
     \caption{$K$ band evolution of the collapse and the accretion
     models. First column: collapse model, face-on, second column:
     accretion model, face-on, third column: collapse model, edge-on,
     fourth column: accretion model, edge-on. The wavelength ranges
     that are shifted into the $K$ band here correspond to ''bands''
     with effective wavelengths (from top to bottom) $613$~nm (between
     $V$ and $R$), $771$~nm ($\sim I$), $1060$~nm (between $I$ and
     $J$), $1470$~nm (between $J$ and $H$), and $2190$~nm ($K$ band).}
     \label{bild07}
   \end{figure*}

   A sample of the synthesized $U$, $V$, and $K$-band face-on and
   edge-on images are plotted in Figs.~\ref{bild05}, \ref{bild06}, and
   \ref{bild07}. Each figure shows the evolution in one colour band
   ($U$, $V$, and $K$) of the collapse and the accretion models. They
   show a time sequence starting at $2.5$~Gyr with the last image
   representing the galaxy at the present epoch (ages: $2.5$, $3.5$,
   $5.5$, $8.5$, and $13.5$~Gyr, corresponding to $z=2.57$, $1.84$,
   $1.07$, $0.49$, and $0.00$). The galaxies are appropriately
   redshifted, but placed at the same distance (something that of
   course cannot be observed, but is necessary to plot all diagrams
   with the same scaling), so one sees in these images the effect of
   the K-correction (the dimming due to redshift and time dilatation),
   but not the dimming due to the increasing distance modulus. For
   each colour, all images are scaled in the same way, such that the
   brightness range spreads $5^{m}$ with the brightest pixel of the
   whole sample overexposed by $3^{m}$ (i.e. the brightest $3^{m}$ are
   plotted in white). Already from these images, one can observe
   interesting evolutionary features. The collapse model shows its
   most interesting features at the beginning of its evolution, when
   its SF is strongest. At a universe time of $3.5$~Gyr, when the core
   is already burnt out, a ring-shaped star forming region appears,
   and collapses at $4.5$~Gyr to a bar and two spiral arms. The bar
   survives until around $10.5$~Gyr, and from there on the galaxy
   appears as an early disk-type galaxy until the present day.

\begin{figure*}
     \includegraphics[width=\textwidth]{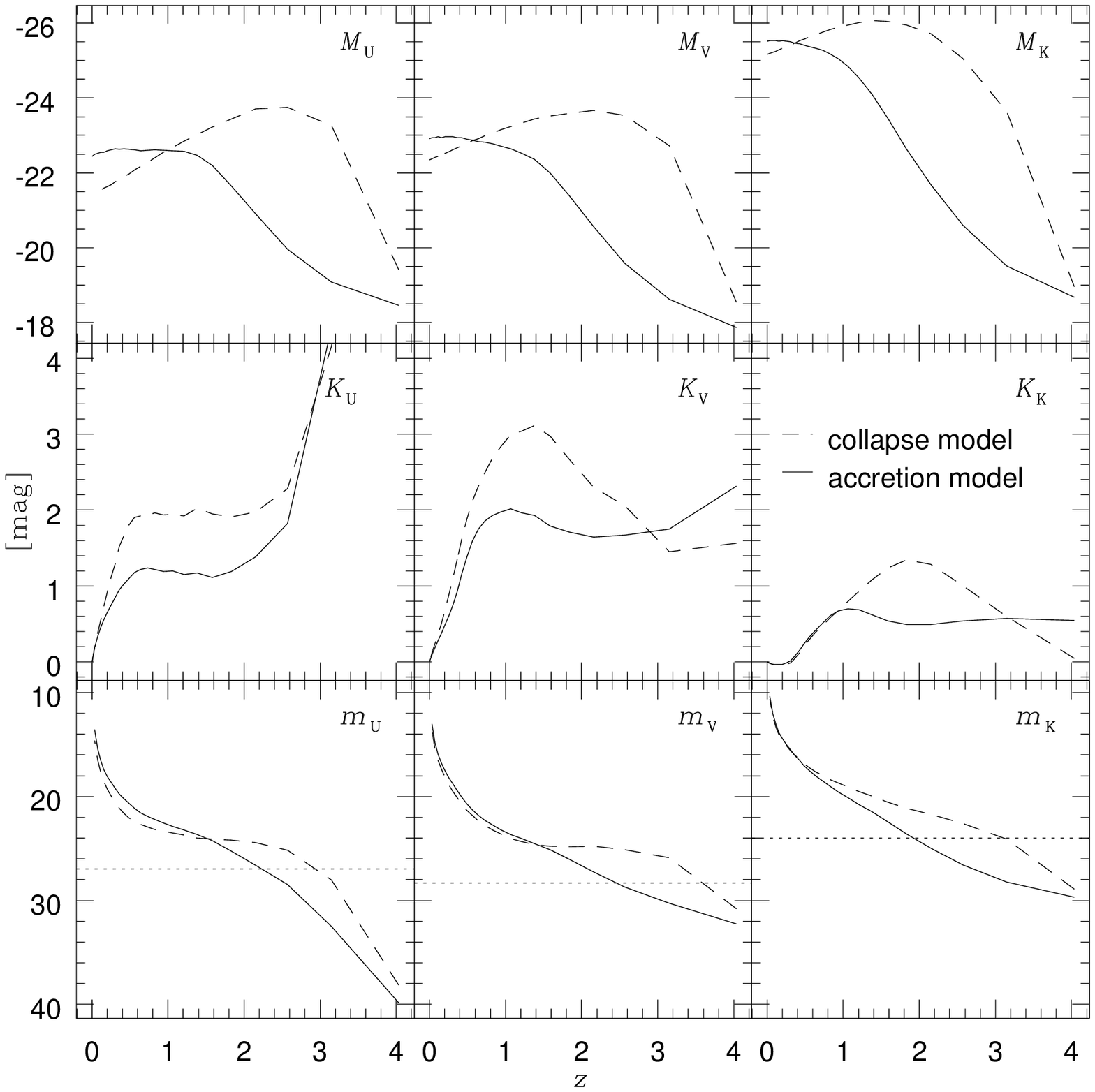}
     \caption{Evolution of integrated photometric properties (from top
     to bottom: absolute magnitudes, K corrections and apparent
     magnitudes) of the accretion model (solid) vs. the collapse model
     (dashed) as a function of redshift in three passbands (from left
     to right: $U$, $V$, and $K$) and at inclination $60^{\circ}$. The
     dotted lines in the apparent magnitude diagrams represent the
     approximate limiting magnitudes in the $F300W$, $F606W$ and
     $F222M$ bands for the Hubble Deep Field North
     \citep{williams_00}. The conversion from evolution time to
     redshift assumes a standard $\Lambda$CDM cosmology.}
     \label{bild08}
   \end{figure*}

   The fact that in the $U$ and $V$ band figures, the $z=0$ images are
   the brightest, whereas these magnitudes are expected to peak at $z
   \simeq 1-2$, when the SFR peaks, is due to the K corrections (see
   the middle row of panels in Fig.~\ref{bild08}), and is not seen in
   the evolution of the intrinsic integrated magnitudes $M_{U}$ and
   $M_{V}$ (Fig.~\ref{bild08}, top row). In the $K$ band, the K
   correction (Fig.~\ref{bild08}, middle right panel) is much smaller
   than in the visual and ultraviolet, which makes the well-known
   property of the $K$ magnitude as a (stellar) mass tracer hold
   approximately true even at high redshift. Therefore, the brightness
   of these images levels off at $z \simeq 1$, when the bulk of the SF
   is completed.  The accretion model shows its most prominent
   features at low redshift. Among them are a bar, formed at $\sim
   6$~Gyr and spiral arms, formed a bit later. Both features survive
   until the present epoch.

\subsection{Integrated photometric quantities}

   In tables~\ref{liste01} to \ref{liste10}, the integrated
   photometric properties (rest frame absolute colours and magnitudes,
   K corrections, and apparent colours and magnitudes according to the
   used cosmology) are summarized for both models in the inclined view
   ($60^{\circ}$). The integrated (intrinsic and apparent) magnitudes
   of the inclined view lie somewhere between the ones of the face-on
   and of the edge-on view, but closer to the magnitudes of the galaxy
   seen face-on. In the accretion model, the inclination does not play
   a role for the intrinsic apparent magnitudes until a redshift of
   $\sim 1.4$, because only then, the disk begins to form. From then
   on, the face-on view becomes gradually brighter than the edge-on
   view in all passbands. At $z=0$ the difference amounts to $\sim
   0.6^{m}$ in $m_{U}$, $m_{B}$, $m_{V}$, $m_{R}$ and $m_{I}$ and
   $\sim 0.15^{m}$ in $m_{K}$. The spatial resolution of the
   simulations is still too low, to resolve the vertical
   stratification of the gaseous disk. Therefore, the absorption in
   the edge-on view is overestimated and the differences between
   face-on and edge-on view are only upper limits.

   The same is seen for the collapse model, but much earlier and with
   a higher magnitude. The two models diverge already at a redshift of
   $\sim 2$ and reach differences of $\sim 1^{m}$ ($m_{U}$, $m_{B}$,
   $m_{V}$, $m_{R}$, $m_{I}$) resp. $\sim 0.5^{m}$ ($m_{K}$).

\begin{figure*}
     \includegraphics[width=\textwidth]{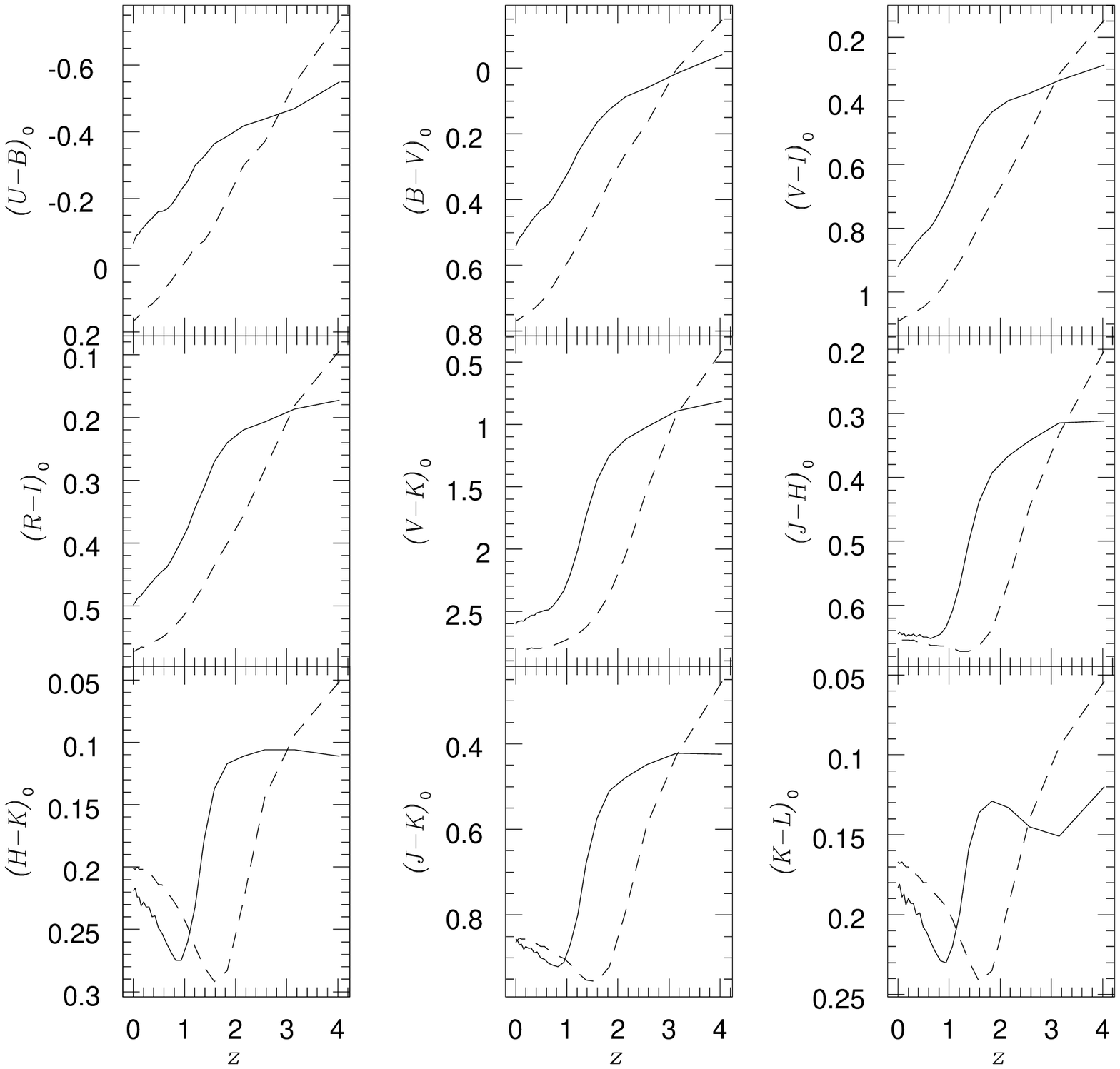}
     \caption{$UBVRIJHKL$ intrinsic integrated colour evolution of the
     accretion (solid) and the collapse model (dashed) as a function
     of redshift and at inclination $60^{\circ}$.}
     \label{bild09}
   \end{figure*}

\begin{figure*}
     \includegraphics[width=\textwidth]{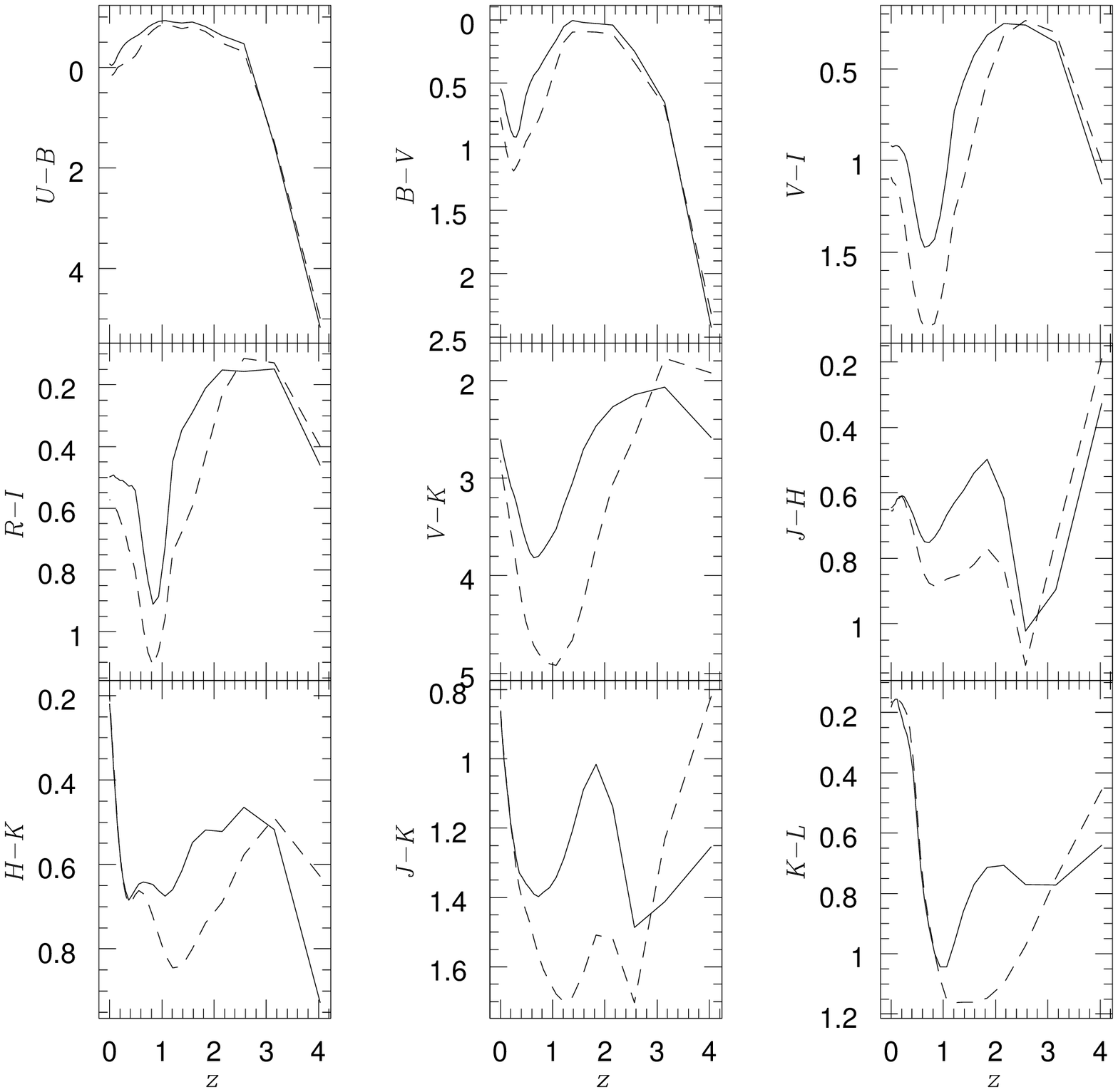}
     \caption{Predicted $UBVRIJHKL$ observed integrated colour
     evolution of the accretion (solid) and the collapse model
     (dashed) as a function of redshift and at inclination
     $60^{\circ}$.}
     \label{bild10}
   \end{figure*}

   Integrated photometric properties (in the inclined view) are shown
   as a function of redshift in Figs.~\ref{bild08} to \ref{bild10}. In
   the absolute magnitude diagrams (Fig.~\ref{bild08}, top row), we
   see again that the brightening of the accretion model begins to
   level off at $z \sim 1$, while the collapse model has already
   passed its zenith. The absolute magnitude evolution is shown here
   only for $M_{U}$, $M_{V}$, and $M_{K}$, but it looks similar in all
   bands.

   The K corrections are shown in the middle row of
   Fig.~\ref{bild08}. The strong $K_{U}$ correction for $z>3$ is
   caused by the Lyman line blanketing and Lyman continuum absorption
   The $K_{B}$ looks similar to $K_{U}$; the same holds for ($K_{V}$,
   $K_{R}$ and $K_{I}$), and for ($K_{K}$, $K_{J}$, $K_{H}$, and
   $K_{L}$).

   In the $m_{U}$, and $m_{V}$ diagrams of Fig.~\ref{bild08}, bottom
   row (here again, the same diagrams for $m_{B}$, $m_{R}$, $m_{I}$,
   $m_{J}$, $m_{H}$, and $m_{L}$ would look very similar), the
   approximate HDF limiting magnitudes of the $F300W$, $F606W$ and
   $F222M$ bands for the Hubble Deep Field North \citep{williams_00}
   are shown as dotted lines. These limits indicate that we should see
   galaxies like the one modelled in the accretion model out to a
   redshift of $2.4$ in the visible and infrared passbands according
   to the $J_{110}$ and $H_{160}$ limits of the NICMOS HDF counterpart
   \citep{thompson_99}.  The HDF objects seen at higher redshift are
   probably the progenitors of more massive early type galaxies (E or
   S0), rather than young disk galaxies. Morphologically, they may be
   hard to distinguish at these redshifts, as the model galaxies do
   not show their disk structure yet. This is confirmed at least
   qualitatively by \citet{abraham_99a, abraham_99b}. The collapse
   model should be seen in the HDF out till $z \simeq 3$ in $m_{U}$ or
   even $\sim 3.5$ in $m_{V}$, $m_{I}$, $m_{J}$ or $m_{H}$. The HDF
   limits also indicate that we need to go $\sim 4^{m}$ fainter in $V$
   and $I$, $\sim 10^{m}$ in $U$ to catch a galaxies like the
   accretion model near birth.

   The intrinsic colours (Fig.~\ref{bild09}) become redder with time
   for both models, with the collapse model colours starting off bluer
   than the accretion model colours, but becoming redder than the
   latter at $z \simeq 3$ in all colours, which is not surprising, if
   we take the star formation history into account.

   The oscillating evolution of the apparent colours
   (Fig.~\ref{bild10}) is a combination of the evolution of the
   corresponding intrinsic colour and the K corrections (see
   Fig.~\ref{bild08}) of the two involved passbands. Again, the
   collapse model starts off bluer in all colours and turns redder
   than the accretion model at $z \simeq 3$. Interestingly though the
   collapse model does not show the very red colours at intermediate
   redshift that are predicted \citep{zepf_97, barger_99} by
   monolithic models like those of \citet{arimoto_87} or
   \citet{matteucci_87}. \citet{zepf_97} predicts, for example, $V-K$
   colours of up to $\sim 7$, whereas our collapse model does reach
   only $V-K=5$, due to the modest, but in the integrated light
   important SF that continues until the present epoch
   (Fig.~\ref{bild02}). This shows that the lack of observed red
   galaxies does not exclude the possibility that some galaxies formed
   early in a single collapse out of one protogalactic gas cloud,
   provided that a minimum SF is maintained after the main
   ''starburst''.

\subsection{Interpretation of the integrated photometric quantities}

\begin{figure*}
     \includegraphics[width=\textwidth]{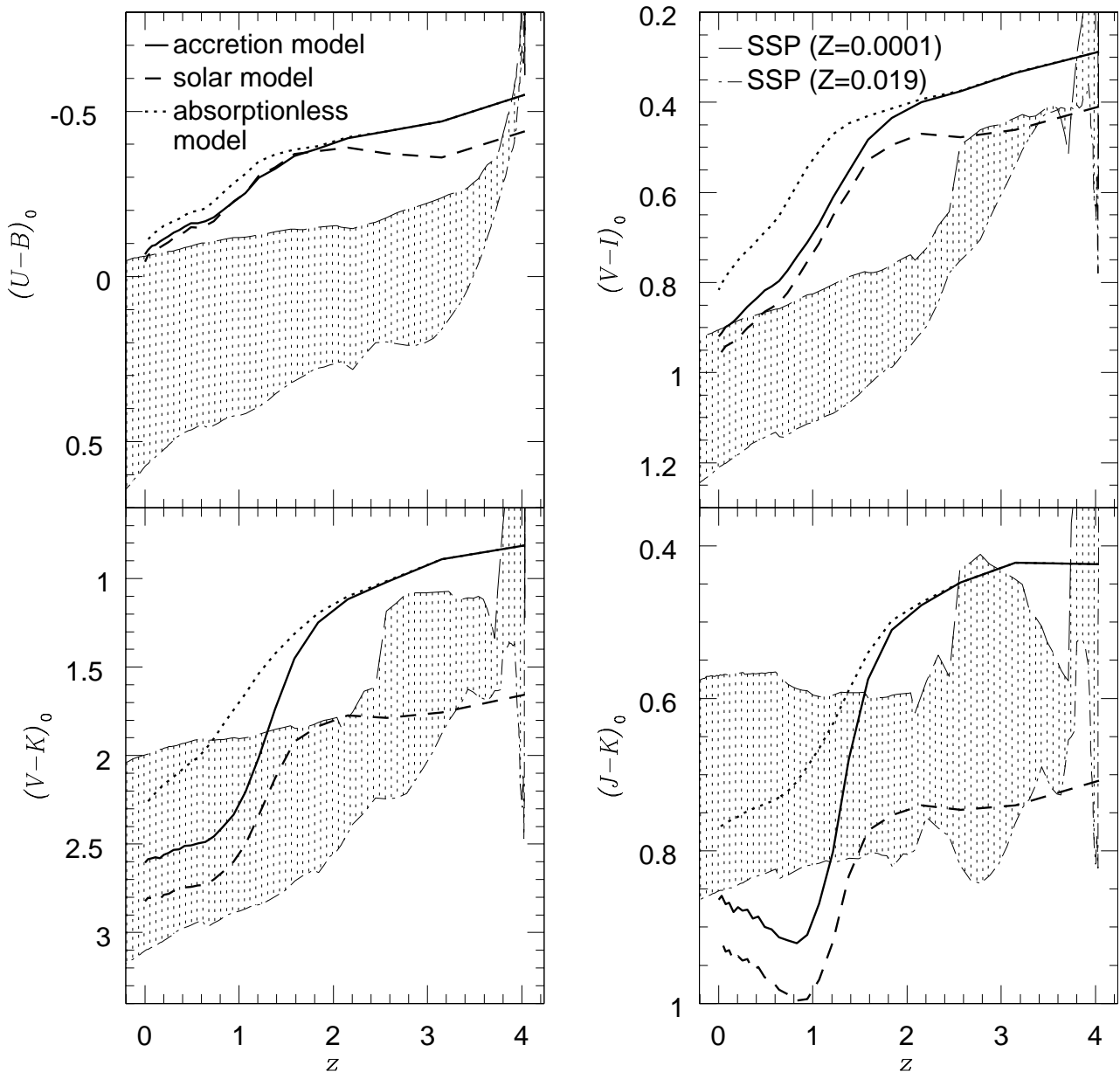}
     \caption{Integrated intrinsic colour evolution in four colours
     ($U-B$, $B-V$, $V-K$, and $J-K$) as function of redshift for the
     accretion model (solid), compared to the evolution of single
     stellar populations, born at $z=4$, with two different
     metallicities (dash-dotted: solar, long dashed: ${\rm [Fe/H]} =
     -2.25$, the region between these two curves is shaded). In
     dashes, the colours are shown for the solar metallicity model,
     and in dotted lines for the absorptionless model.}
     \label{bild11}
   \end{figure*}

   In the following, we concentrate on the accretion model. In order
   to interpret the evolution of the intrinsic colours
   (table~\ref{liste01} and \ref{liste02}), we compare them with the
   colour evolution of SSPs (Fig.~\ref{bild11}). These were also
   produced with the GISSEL code, using the same tracks, stellar
   library and IMF as for the galaxy model, so any systematics
   stemming from the input SSP spectra should cancel out (actually
   these are the SSPs, from which the model galaxy colours were
   derived). The two SSPs shown here are for solar metallicity
   (dash-dotted), which is not far above the average metallicity at
   which the galaxy model ends, and ${\rm [Fe/H]} = -2.252$ (long
   dashes), the lowest metallicity available. The regions between
   these two curves are shaded to show the ranges in which the colours
   of SSPs evolve. The solid lines represent the colour evolution of
   the accretion model. In order to identify metallicity and
   absorption effects, the solar metallicity and the absorptionless
   models (see Section~\ref{chapter3}) are shown as dashed
   resp. dotted lines.

   Of course all three, the accretion model, the solar metallicity
   model, and the absorptionless models behave smoother with redshift
   than the SSPs, because they represent convolutions of SSPs with an
   SFR. In $(U-B)_{0}$ (the metallicity indicator for individual
   stars), $(V-I)_{0}$, and other ultraviolet and optical colours not
   shown here, ongoing star formation dominates the colours, so the
   models remain even bluer than for the ${\rm [Fe/H]} = -2.252$ SSP
   during the entire evolution. The SF determines the colour evolution
   as long as SF continues and metallicity and absorption change these
   two colours by less than $0.1^{m}$. It is expected that the colours
   will rapidly (within a few Gyr) tend towards the lower (solar
   metallicity) curve once SF is completed. In the infrared colours
   $(V-K)_{0}$ and $(J-K)_{0}$, where SF does not leave such a strong
   imprint, we see a combination of SF and internal reddening, placing
   the colours well below the ${\rm [Fe/H]} = -2.252$ SSP evolution
   from $z\simeq 1.4$ on. $(J-K)_{0}$ comes out even redder than the
   solar metallicity SSP for a few Gyr around $z=1$, when the gas
   density in the centre of the model galaxy is the highest. The fact
   that the absorptionless model (dotted) lies between the two SSPs
   proves that this is an absorption effect. These colours are of
   course also expected to tend towards the ones of the solar SSP with
   time. As can be seen from the colour evolution of the solar
   metallicity model (dashed), the metallicity effect is relatively
   minor for $z\stackrel{_<}{_\sim} 1.4$. With $\simeq 0.2^{m}$ in
   $(V-K)_{0}$ and $\simeq 0.1^{m}$ in $(J-K)_{0}$, it is stronger
   than in the ultraviolet and optical colours, even though these
   infrared colours are known to be metallicity insensitive for
   individual stars.

   Obviously, different rules apply for the metallicity dependence of
   colours for composite stellar populations than for individual
   stars. This is explained by the fact that, by comparing populations
   of different metallicities, we are not looking at stars of the same
   stellar parameters, but at stars of the same age. Populations of
   the same age do not necessarily show the same metallicity
   dependence as stars of the same stellar parameters (that is, a
   strong metallicity dependence in the ultraviolet and metallicity
   independence in the infrared). As stars of different ${\rm [Fe/H]}$
   develop differently (metal-rich stars have longer lifetimes than
   metal-poor stars of the same mass), the infrared colours do depend
   on metallicity for stellar populations. This can actually already
   be seen from the colour evolutions of SSPs in
   Fig.~\ref{bild11}. Hence, in our galaxy model, metallicity effects
   can be observed in $(J-K)_{0}$, whereas in $(U-B)_{0}$, they are
   suppressed as long as SF continues.

   \begin{figure*}
     \includegraphics[width=\textwidth]{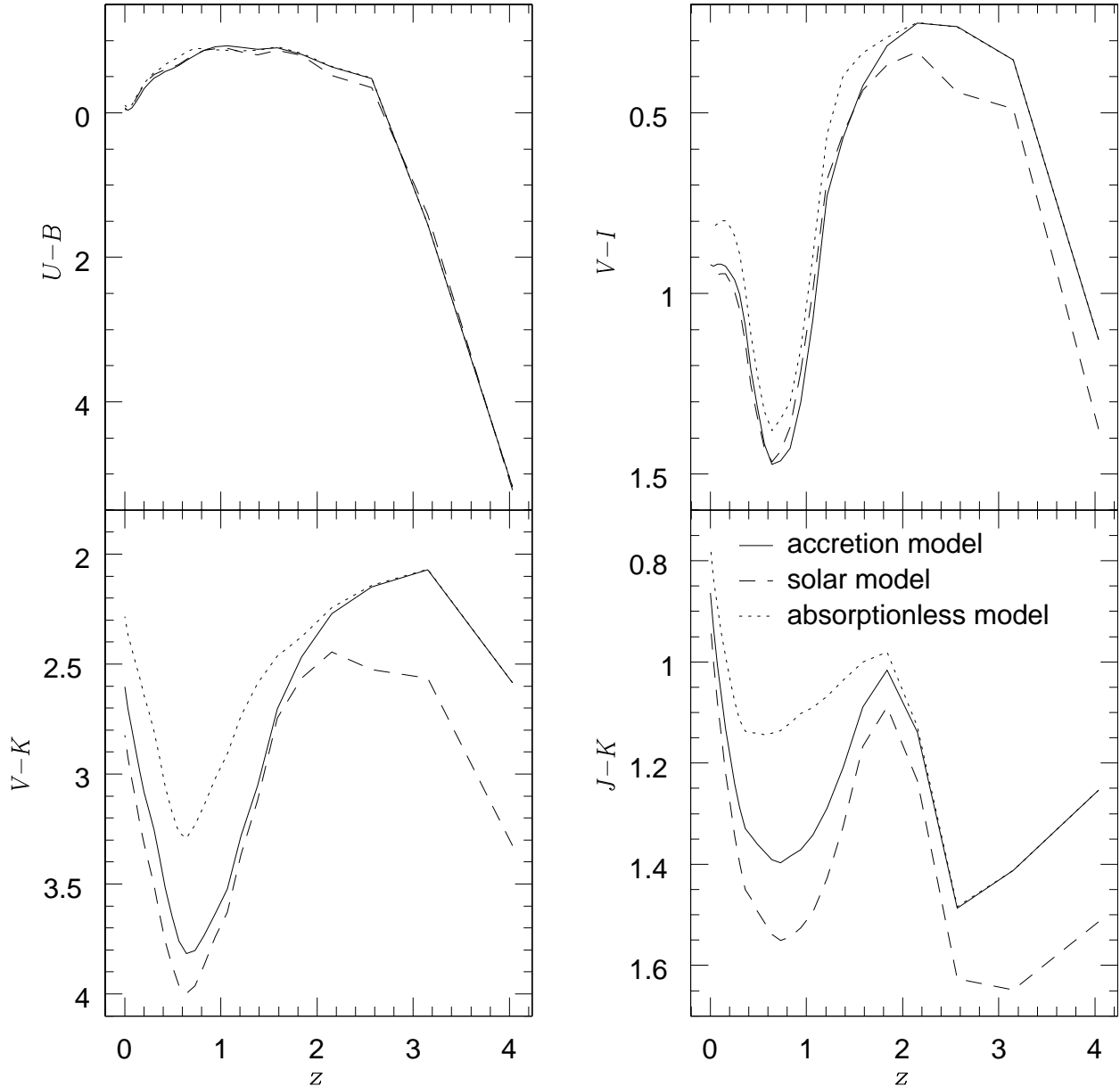}
     \caption{Integrated observed colour evolution in four colours
     ($U-B$, $B-V$, $V-K$, and $J-K$) ,as function of redshift, for
     the accretion model (solid) compared to the solar metallicity
     model (dashed), and the absorptionless model (dotted).}
     \label{bild12}
   \end{figure*} 

   So far, we have discussed only intrinsic colours. For the
   comparison with real galaxies, we have to look at the evolution of
   the redshifted intrinsic colours. This evolution is shown in
   Fig.~\ref{bild12} for the same colours as in Fig.~\ref{bild11} (the
   comparison with SSPs is omitted here). Again, $U-B$ is only
   slightly affected by both metallicity and absorption, and mainly
   reflects SF and the absorption of intervening gas at high
   redshifts. In $V-I$, the differences between the accretion model
   and the solar metallicity model amount to $0.2^{m}$ at high
   redshift, but become negligible at $z \simeq 2$. From then on,
   absorption effects become important, but they do not change $V-I$
   by more than $0.15^{m}$. Metallicity is crucial for the evolution
   of the infrared colours $V-K$ and $J-K$. At high redshift, it can
   change $V-K$ by up to $0.75^{m}$, and $J-K$ by $0.25^{m}$. At
   $z=0$, the difference is still $0.2^{m}$ resp. $0.05^{m}$.

   Absorption is most important at $z \stackrel{_<}{_\sim} 2$. At $z
   \simeq 1$, it changes $V-K$ by $0.6^{m}$, and $J-K$ by
   $0.25^{m}$. The main difference between the evolution of the two
   infrared colours shown here is seen from $z \simeq 1.8$ to $z
   \simeq 2.6$, when $J-K$ becomes bluer by $\sim 0.4^{m}$, due to the
   $4000$~\AA~break that wanders through the $J$ band between these
   redshifts.

   The lack of absorption effects in all colours at high redshift in
   these models is explained by the lack of gas in the dark halo at
   this epoch. Of course, the gas that falls in at a later stage is
   already there outside the halo, and its absorption would in
   principle have to be included in our models too, but it is
   distributed over such a large volume that only a negligible
   fraction will be located in the line of sight.

   From the large metallicity and absorption influences on colours, it
   follows that the metallicity distribution of the stars and the
   internal absorption by gas must be taken into account when deriving
   colours from galaxy models.

\subsection{Bulge colours}

\begin{figure}
     \includegraphics[width=\columnwidth]{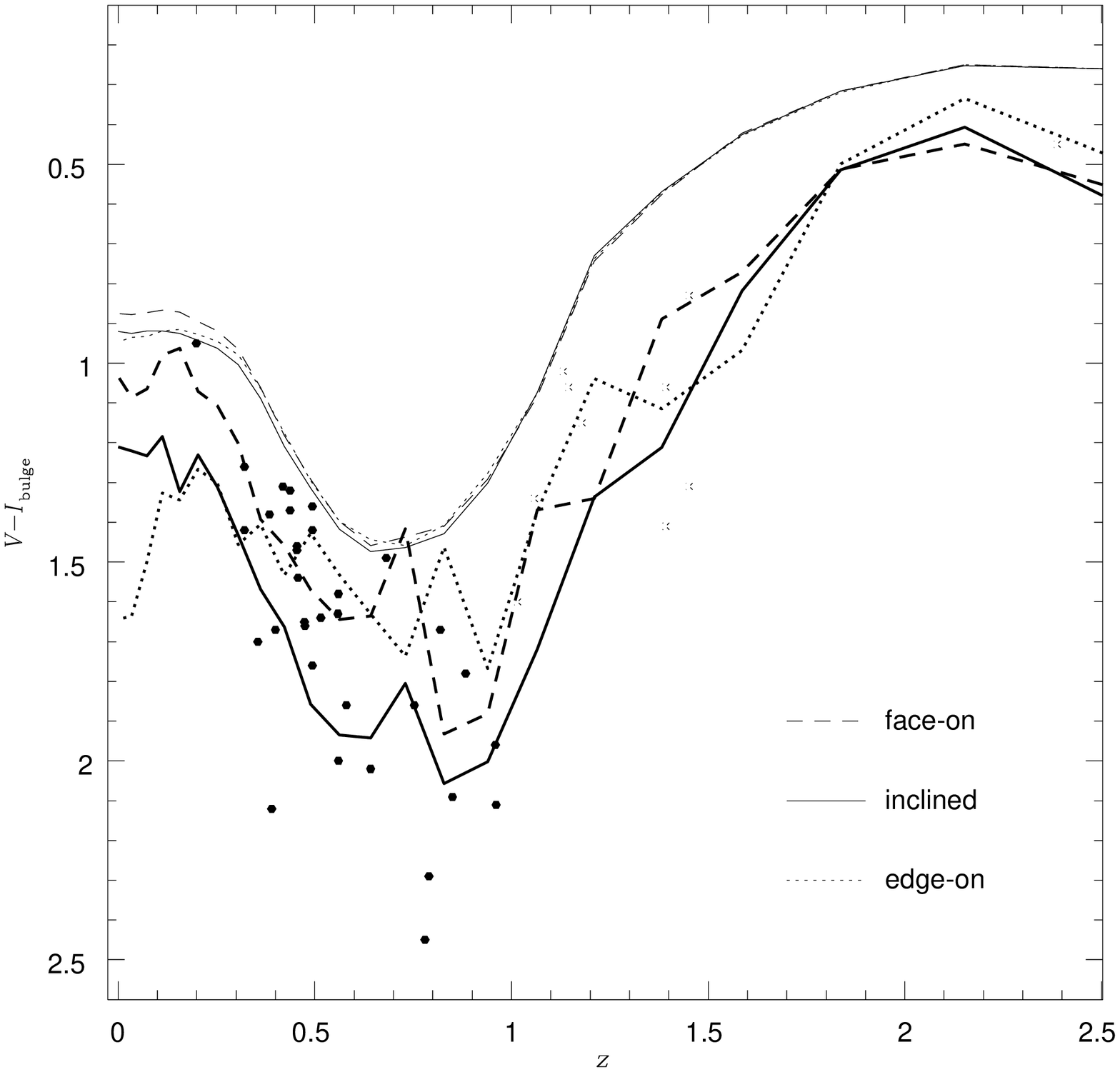}
     \caption{Predicted bulge $V-I$ colours (thick lines) as function
     of redshift from the accretion model for three different
     inclinations (dashed: face-on, solid: inclined by $60^{\circ}$,
     dotted: edge-on), compared to HDF disk galaxy bulge colours
     \citep{ellis_01}, transformed into the standard system according
     to \citet{fukugita_95} (thick points). For the crosses ($z>1$),
     the colours were kept in the HST $V_{606}-I_{814}$ system, as no
     transformation equation was available. The integrated model
     colours for the three inclinations are shown as thin lines.}
     \label{bild13}
   \end{figure}

\begin{figure}
     \includegraphics[width=\columnwidth]{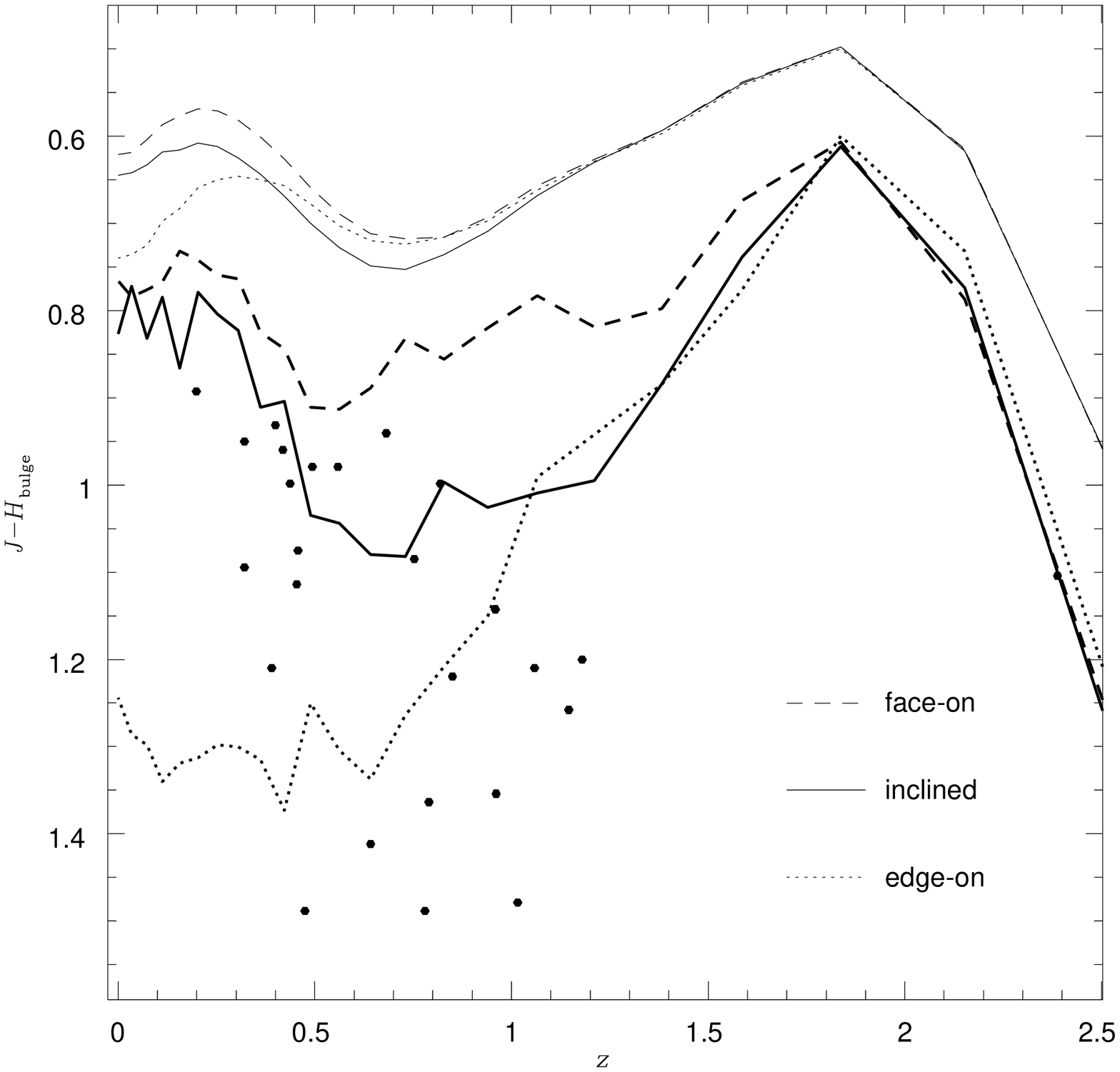}
     \caption{Predicted bulge $J-H$ colours (thick lines) as function
     of redshift from the accretion model for three different
     inclinations (dashed: face-on, solid: inclined by $60^{\circ}$,
     dotted: edge-on), compared to HDF disk galaxy bulge colours
     \citep{ellis_01}, transformed into the standard system according
     to \citet{stephens_00} (thick points). The integrated model
     colours for the three inclinations are shown as thin lines.}
     \label{bild14}
   \end{figure}

   To test the properties of our models, we compared the colours of
   HDF-N disk galaxy bulges from \citet{ellis_01} with our results. We
   expect bulge colours to be more accurate than integrated galaxy
   colours, as they are usually measured within isophotes well above
   the noise level. To derive bulge colours from our models, we
   located for each ''frame'' the highest concentration of light, and
   calculated the integrated colours over a range of $1.25$~kpc around
   this centre (corresponding to around $20$ ''pixels''). Varying this
   ''aperture size'' showed that this is a reasonable value. It also
   corresponds well to the aperture used by
   \citet{ellis_01}. Tables~\ref{liste11} and \ref{liste12} summarize
   our bulge colours for the face-on and the edge-on view, and in
   Figs.~\ref{bild13} and \ref{bild14}, the redshift evolutions of
   $V-I$ and $J-H$ are plotted, as well as the empirical data, which
   had to be transformed from the HST $V_{606}-I_{814}$
   resp. $J_{110}-H_{160}$ system into Johnson-Cousins $V-I$
   \citep{fukugita_95} and $J-H$ \citep{stephens_00}. The thick points
   in Figs.~\ref{bild13} and \ref{bild14} show the transformed data,
   while the crosses in the $V-I$ diagram are still in the HST system,
   as no transformation was available. The model predictions for the
   face-on (dashed), the inclined (solid), and edge-on (dotted) view
   are drawn as thick lines, and for comparison, the integrated
   colours of the same models are shown as thin lines. The
   \citet{ellis_01} integrated galaxy colours are not shown here in
   order not to overload the figure. On average, they are only around
   $0.1^{m}$ bluer than their bulge colours, whereas our model
   predicts them to be $0.2^{m}$ to $0.3^{m}$ bluer. This is probably
   due to the fact, that we calculate the model galaxy colours by
   using all the light out to $40$~kpc from the galactic centre. The
   model shows a colour gradient in the sense that the galaxy is bluer
   in the outer parts. In fact, the colour of the model galaxy
   integrated over the inner $10$~kpc is less than $0.1^{m}$ bluer
   than the bulge colour. This is why, in our comparison with
   observations, we use bulge colours rather than full galaxy
   colours. Clearly, the models reproduce the observed colours
   well. One can argue that our $J-H$ predictions are too blue, but
   they are within the measuring errors ($\sim 0.15^{m}$).

   The different evolutions of the edge-on model and the face-on model
   bring us to an interesting result: At low redshift, the models
   predict much redder bulge colours for edge-on than for face-on
   galaxies. The inclined model follows the face-on model more or
   less. The difference amounts to $0.5^{m}$ in both $V-I$ and $J-H$
   (the two colours studied here, but the effect is present in all
   colours). This is due to absorption from the dust in the disk. In
   $J-H$, this absorption begins to take its toll already at a
   redshift of $1.2$, whereas in $V-I$, it does not appear until
   $z\simeq 0.4$. This corresponds well with the fact that the $V$ and
   $I$ band are shifted into the $J$ and $H$ band at redshift $\sim
   1$. In the edge-on view, the absorption effect is overestimated as
   a result of the limited spatial resolution in our models. In the
   models, a significant fraction of the bulge is reddened by the
   thick gas disk, while in observed galaxies, the absorbing material
   is concentrated in a thin disk, which affects only a small fraction
   of the bulge. Nevertheless, this result shows that it is important
   to know the inclination of disk galaxies, in order to interpret
   their bulges' colours, especially if one wants to compare the
   colour evolution of these bulges with the colour evolution of
   elliptical galaxies.

\subsection{Magnitude - and colour profiles}
\label{chapter4.4}

   Finally, we study the profiles in different colours and magnitudes
   and their possible correlations with the profiles of the physical
   quantities presented in Section~\ref{chapter2}, Fig.~\ref{bild03},
   stellar mass surface density, stellar particle age, and stellar
   metallicity.

   \begin{figure*}
     \includegraphics[width=\textwidth]{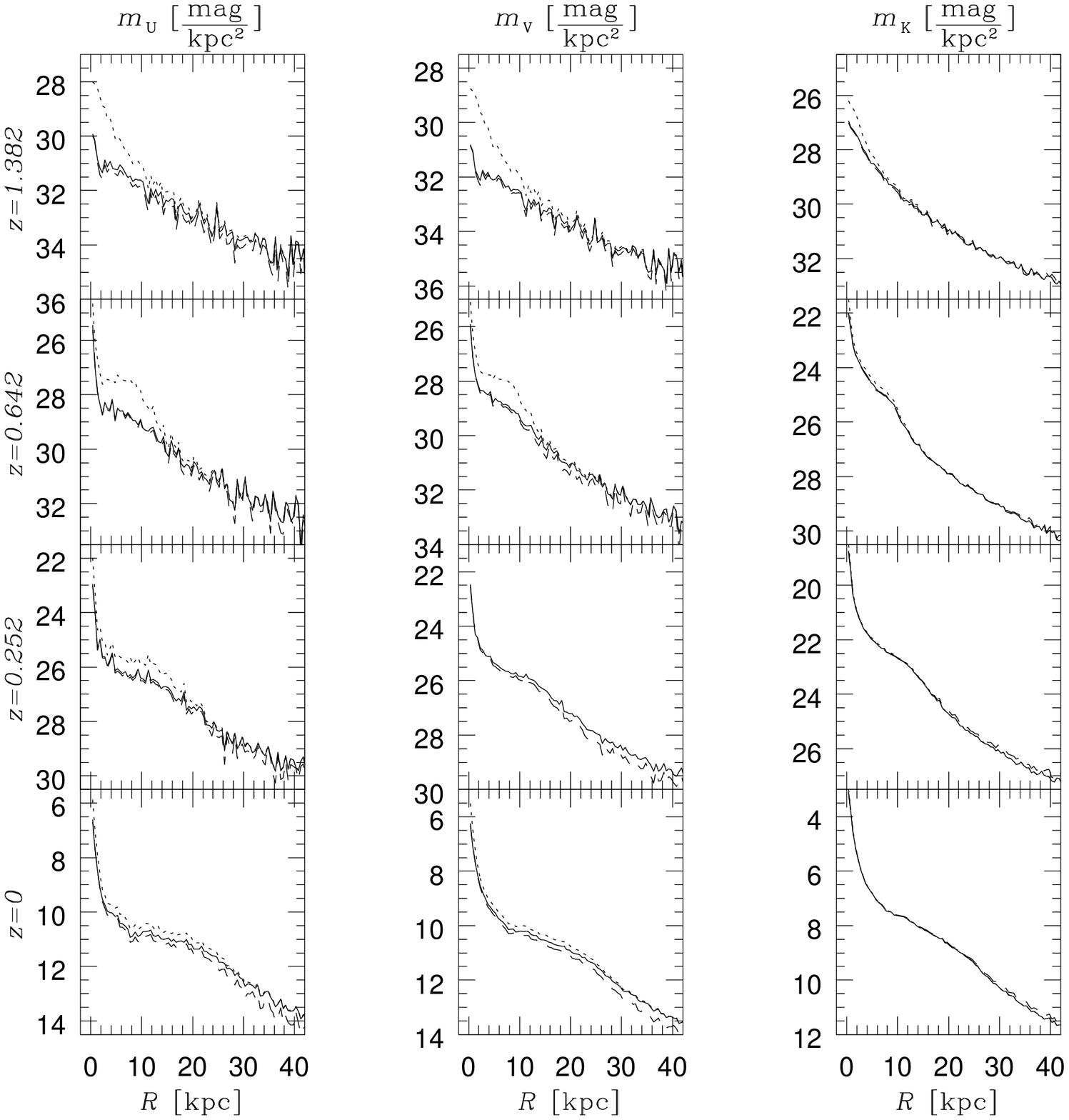}
     \caption{Apparent magnitude profiles in the $U$, $V$, and $K$
     (from left to right) passbands for the accretion model (solid),
     compared with the profiles of the solar metallicity model
     (dashed) and the absorptionless model (dotted), for the same
     redshifts as in Fig.~\ref{bild03}.}
     \label{bild15}
   \end{figure*} 

   \begin{figure*}
     \includegraphics[width=\textwidth]{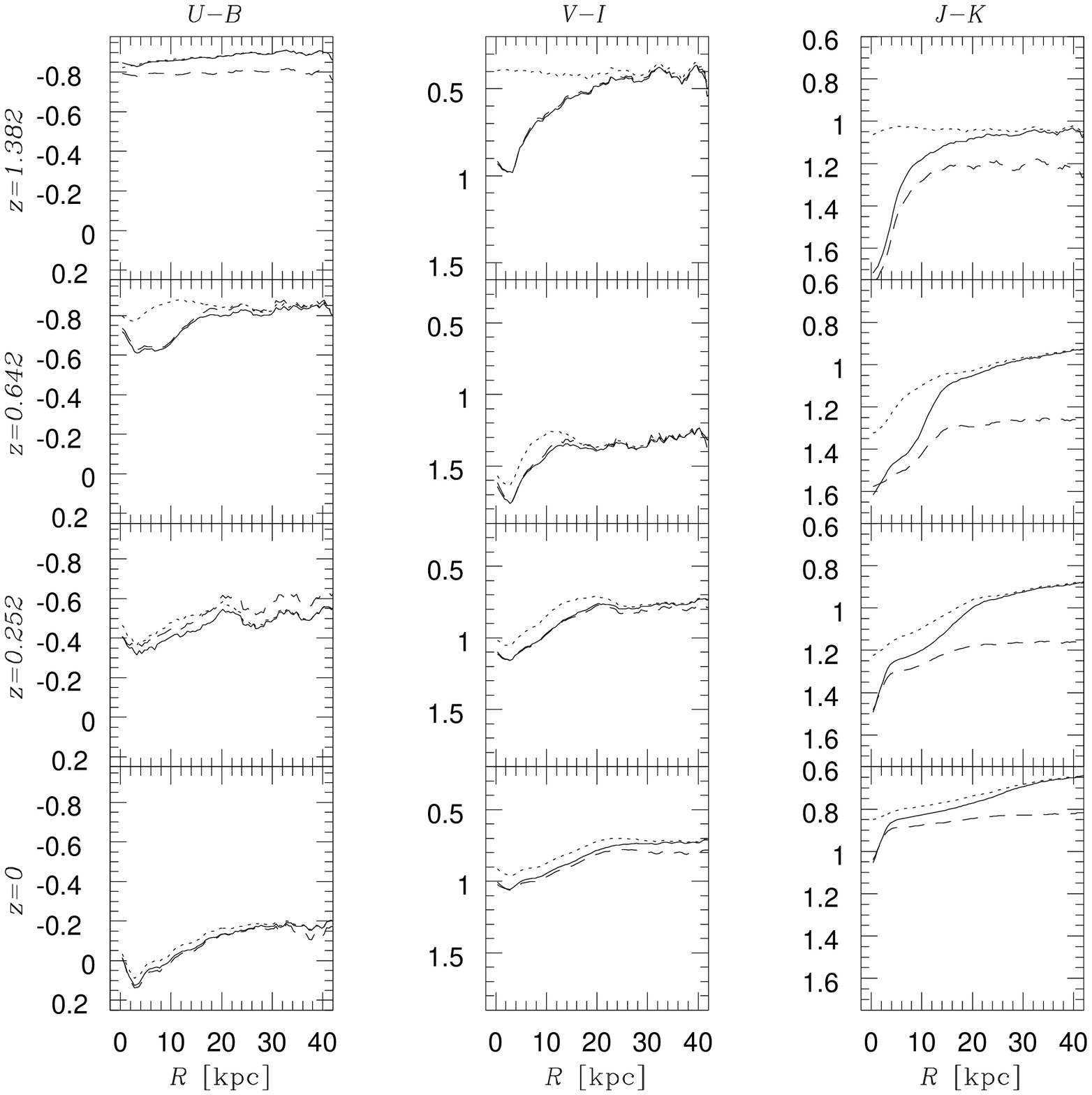}
     \caption{Colour profiles in $U-B$, $V-I$, and $J-K$ (from left to
     right) for the accretion model (solid), compared with the solar
     metallicity model (dashed) and the absorptionless model (dotted),
     for the same redshifts as in Fig.~\ref{bild03}.}
     \label{bild16}
   \end{figure*} 

   The apparent magnitude profiles (calculated in the same way as for
   the quantities in Fig.~\ref{bild03}) in $U$, $V$, and $K$ are shown
   in Fig.~\ref{bild15} as solid lines. In order to see if there are
   any metallicity or absorption effects on these profiles, the same
   profiles are drawn as dashed lines for the solar metallicity model,
   and as dotted lines for the absorptionless model. All magnitude
   profile plots are scaled in the same way which means $1^{m}$ has
   the same size in all plots, so they can directly be compared.

   As expected, all three profiles reflect the mass density, whereas
   the metallicity gradient causes only a small modification, which in
   real data can hardly be disentangled from the mass density
   contribution. The best mass density tracer is obviously $m_{K}$,
   which is well known for this property. It follows the mass density
   perfectly, and shows almost no metallicity or absorption
   influence. An agreeable property of $m_{K}$ as a mass tracer is
   that it holds true even for the redshifted models, although we are
   actually looking at wavelength regions corresponding to the $I$,
   $J$, and $H$ bands there. In $m_{U}$ and $m_{V}$, absorption
   amounts to $\sim 2^{m}$ in the centre at redshifts around 1.

\begin{figure}
     \includegraphics[width=\columnwidth]{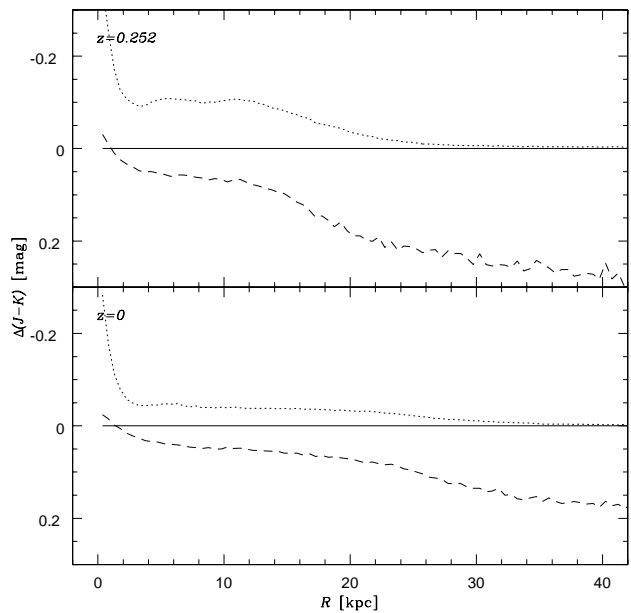}
     \caption{Differences in the $J-K$ radial profiles between the
     solar metallicity model (dashed) resp. the absorptionless model
     (dotted) and the regular accretion model from
     Fig.~\ref{bild16}, in the sense solar metallicity
     (absorptionless) $-$ regular model. The solid line represents
     zero difference.}
     \label{bild17}
   \end{figure}

   More surprising are the results for the colour gradients. In
   Fig.~\ref{bild16}, they are shown for $U-B$, $V-I$, and $J-K$ as
   solid lines vs. the same profiles for the solar metallicity model
   (dashed) and the absorptionless model (dotted). Again, they are
   plotted on the same scale for each redshift. Surprisingly, $U-B$
   and $V-I$ prove almost metallicity independent (the $U-B$ profile
   at $z=1.382$ should not be over-interpreted, because the stellar
   library was not calibrated in the range that is shifted into the
   $U$ and $B$ here), whereas metallicity seems to leave a stronger
   effect on $J-K$ (at the same time, absorption is negligible here),
   a result we already encountered for the time evolution of the
   integrated light of the two models. Indeed, the profiles of the
   differences $\Delta(J-K)$ between the regular accretion model and
   the solar metallicity model (shown for low redshifts in
   Fig.~\ref{bild17}) are well correlated with the ${\rm [Fe/H]}$
   profile (Fig.~\ref{bild03}). There is some profit to be taken from
   this, due to the fact that the solar metallicity model profile is
   more or less horizontal (apart from the inner bulge;
   Fig.~\ref{bild16}), at least at low redshift. This means that a
   $J-K$ gradient should be directly related to the metallicity
   gradient, even though the absolute colour will not necessarily
   yield the absolute ${\rm [Fe/H]}$ value. We find
   \begin{equation}
     \frac{d{\rm [Fe/H]}}{dR} \simeq 4\cdot\frac{d(J-K)}{dR},
     \hspace{2cm} z \leq 0.5.
   \end{equation}
   Of course, this finding needs to be confirmed observationally, but
   it is an indication that $J-K$ could prove a useful metallicity
   gradient tracer.

\section{Discussion and Conclusions}

   We present the spectral analysis of two chemo-dynamical galaxy
   formation models, evaluated with a state of the art evolutionary
   code and spectral library. The programme transforming the models
   into spectral properties takes into account the three-dimensional
   distribution of the stars and the interstellar matter. It includes
   internal gas absorption and is also able to include foreground
   reddening.

   We obtain two-dimensional $UBVRIJHKLM$ images of the model
   galaxies, giving apparent magnitudes and colours in up to $160
   \times 160$ pixels. We also obtain intrinsic and apparent
   integrated spectra and intrinsic colours of the model galaxies. All
   of these quantities can be calculated with a time resolution of
   $10$~Myr. In the present work, they were calculated in time steps
   of $0.5$~Gyr. We find that
   \begin{enumerate}
     \item The integrated colours of the model galaxy depend more
     strongly on the metallicity of the stars in the infrared
     ($(V-K)_{0}$, $(J-K)_{0}$, $V-K$, $J-K$) than in the optical and
     ultraviolet.
     \item In the ultraviolet ($(U-B)_{0}$, $U-B$), star formation
     blankets metallicity - and absorption effects.
     \item In the infrared and optical, metallicity effects are
     crucial for $z > 1.5$; at lower redshifts absorption becomes more
     important than metallicity. Thus both the metallicity
     distribution of the stars and the internal absorption by gas must
     be taken into account when deriving colours from galaxy models.
     \item At low redshifts, bulge colours depend on the inclination
     of the model galaxy due to absorption from the disk. In our
     model, bulges are up to $0.5^{m}$ redder in edge-on projection
     than when seen face-on, which however is an upper limit because
     of the limited spatial resolution of the underlying numerical
     model.
     \item We confirm the usefulness of $m_{K}$ as a mass tracer, and
     find indications that metallicity gradients manifest themselves
     in $J-K$ (at low redshift), which would give us a means to
     measure ${\rm [Fe/H]}$ gradients from $J-K$ gradients, if
     confirmed observationally.
     \item A comparison of our model colours for disk galaxy bulges
     with empirical data from the HDF North \citep{ellis_01} shows
     good agreement, confirming that the star formation history and
     time dependent gas density distribution of the models are
     realistic.
   \end{enumerate}
   The produced colour images and spectra have not been fully
   exploited yet by far. There is still a lot more information to
   extract and a lot more conclusions to be drawn from these
   files. This will be done in future work.

   Our result that internal absorption is crucial at $z<1.5$ shows
   that it is necessary in any galaxy formation model to have a
   realistic description of the gas component, if galaxy colours are
   to be predicted reliably. This requires at least a 2-phase model of
   the interstellar matter in which a cold star-forming medium
   coexists with a hot component which absorbs most of the energy and
   metal return from massive stars, i.e., a chemo-dynamical approach.
   Three-dimensional and high-resolution chemo-dynamical models, when
   embedded in a realistic cosmological model, allow us not only to
   predict the detailed morphology and colours of forming galaxies,
   but also to investigate the physical processes relevant during the
   formation and evolution of galaxies. Much further work on improving
   the present models is needed, but will be very rewarding.

   The spectro-photometric programme that transforms the quantities
   calculated by the galaxy models into spectral properties has great
   potential, as it calculates colours and spectra in a realistic way
   including, e.g., spatially resolved absorption, using as few
   simplifications and assumptions as possible. Improvements of the
   input ingredients (the stellar evolutionary tracks, the stellar
   library, the absorption law) can easily be implemented. Other
   possible improvements are the inclusion of emission from HII
   regions and planetary nebulae, or the inclusion of supernovae
   spectra. The results do not need to be restricted to the
   $UBVRIJHKLM$ system. Two dimensional distributions can in principle
   be calculated in any colour system, or for other spectral
   properties, such as line strength indices.

\begin{acknowledgements}
   This work was supported by the Swiss National Science Foundation. 
\end{acknowledgements}

\appendix

\section{Intrinsic magnitudes and colours}

\begin{table*}
\begin{center}
\caption{Intrinsic integrated magnitudes and colours of the accretion
model.}
\label{liste01}
$$
\begin{array}{rrrrrrrr}\hline
   z & t [Gyr] & M_{Bol} &  M_{U} &  M_{V} &  M_{K} & (U-B)_{0} &
   (B-V)_{0}\\ \hline
 4.035 &  1.5 & -19.663 & -18.451 & -17.859 & -18.672 & -0.550 & -0.042 \\
 3.150 &  2.0 & -20.155 & -19.077 & -18.624 & -19.516 & -0.469 &  0.016 \\
 2.569 &  2.5 & -21.015 & -19.961 & -19.581 & -20.601 & -0.439 &  0.059 \\
 2.153 &  3.0 & -21.970 & -20.899 & -20.568 & -21.686 & -0.418 &  0.087 \\
 1.837 &  3.5 & -22.663 & -21.646 & -21.385 & -22.634 & -0.387 &  0.126 \\
 1.586 &  4.0 & -23.190 & -22.202 & -22.001 & -23.452 & -0.365 &  0.164 \\
 1.382 &  4.5 & -23.430 & -22.465 & -22.355 & -24.091 & -0.326 &  0.216 \\
 1.211 &  5.0 & -23.568 & -22.571 & -22.528 & -24.537 & -0.299 &  0.256 \\
 1.065 &  5.5 & -23.606 & -22.589 & -22.642 & -24.847 & -0.252 &  0.305 \\
 0.939 &  6.0 & -23.658 & -22.601 & -22.714 & -25.048 & -0.226 &  0.339 \\
 0.828 &  6.5 & -23.696 & -22.612 & -22.781 & -25.185 & -0.200 &  0.369 \\
 0.730 &  7.0 & -23.709 & -22.603 & -22.817 & -25.273 & -0.180 &  0.394 \\
 0.642 &  7.5 & -23.726 & -22.593 & -22.838 & -25.326 & -0.168 &  0.413 \\
 0.562 &  8.0 & -23.755 & -22.616 & -22.879 & -25.374 & -0.161 &  0.424 \\
 0.489 &  8.5 & -23.786 & -22.634 & -22.903 & -25.409 & -0.162 &  0.431 \\
 0.423 &  9.0 & -23.807 & -22.645 & -22.938 & -25.449 & -0.151 &  0.444 \\
 0.362 &  9.5 & -23.797 & -22.625 & -22.938 & -25.469 & -0.141 &  0.454 \\
 0.305 & 10.0 & -23.819 & -22.636 & -22.968 & -25.500 & -0.133 &  0.465 \\
 0.252 & 10.5 & -23.801 & -22.610 & -22.964 & -25.514 & -0.123 &  0.477 \\
 0.203 & 11.0 & -23.802 & -22.594 & -22.968 & -25.529 & -0.114 &  0.488 \\
 0.157 & 11.5 & -23.777 & -22.552 & -22.945 & -25.525 & -0.106 &  0.499 \\
 0.113 & 12.0 & -23.779 & -22.547 & -22.961 & -25.536 & -0.095 &  0.509 \\
 0.073 & 12.5 & -23.760 & -22.522 & -22.945 & -25.526 & -0.092 &  0.515 \\
 0.034 & 13.0 & -23.755 & -22.495 & -22.943 & -25.529 & -0.081 &  0.529 \\
 0.000 & 13.5 & -23.709 & -22.437 & -22.911 & -25.515 & -0.067 &  0.541 \\
 \hline
\end{array}
$$
\end{center}
\end{table*}
 
\begin{table*}
\begin{center}
\caption{Intrinsic integrated magnitudes and colours of the accretion
model (part 2).}
\label{liste02}
$$
\begin{array}{rrrrrrrrr}\hline
  z & (V-I)_{0} & (V-K)_{0} & (R-I)_{0} & (J-H)_{0} & (H-K)_{0} &
  (J-K)_{0} & (K-L)_{0} & (K-M)_{0} \\ \hline
  4.035 & 0.288 & 0.813 & 0.173 & 0.312 & 0.111 & 0.424 & 0.120 & 0.545 \\
  3.150 & 0.335 & 0.892 & 0.187 & 0.315 & 0.106 & 0.422 & 0.151 & 0.582 \\
  2.569 & 0.376 & 1.020 & 0.207 & 0.343 & 0.106 & 0.448 & 0.145 & 0.585 \\
  2.153 & 0.400 & 1.118 & 0.220 & 0.367 & 0.111 & 0.478 & 0.133 & 0.583 \\
  1.837 & 0.435 & 1.249 & 0.240 & 0.393 & 0.117 & 0.510 & 0.129 & 0.598 \\
  1.586 & 0.483 & 1.451 & 0.270 & 0.438 & 0.137 & 0.575 & 0.136 & 0.660 \\
  1.382 & 0.553 & 1.736 & 0.312 & 0.501 & 0.179 & 0.680 & 0.159 & 0.796 \\
  1.211 & 0.611 & 2.009 & 0.344 & 0.568 & 0.232 & 0.801 & 0.199 & 0.976 \\
  1.065 & 0.669 & 2.205 & 0.376 & 0.609 & 0.260 & 0.869 & 0.220 & 1.075 \\
  0.939 & 0.711 & 2.334 & 0.397 & 0.634 & 0.275 & 0.910 & 0.230 & 1.131 \\
  0.828 & 0.743 & 2.404 & 0.413 & 0.645 & 0.275 & 0.921 & 0.229 & 1.139 \\
  0.730 & 0.774 & 2.456 & 0.428 & 0.649 & 0.268 & 0.917 & 0.223 & 1.128 \\
  0.642 & 0.797 & 2.488 & 0.439 & 0.652 & 0.261 & 0.913 & 0.217 & 1.115 \\
  0.562 & 0.809 & 2.495 & 0.445 & 0.650 & 0.253 & 0.903 & 0.211 & 1.095 \\
  0.489 & 0.818 & 2.506 & 0.450 & 0.650 & 0.249 & 0.900 & 0.208 & 1.086 \\
  0.423 & 0.832 & 2.511 & 0.456 & 0.646 & 0.239 & 0.886 & 0.199 & 1.060 \\
  0.362 & 0.843 & 2.531 & 0.462 & 0.649 & 0.240 & 0.889 & 0.200 & 1.065 \\
  0.305 & 0.854 & 2.532 & 0.467 & 0.645 & 0.232 & 0.877 & 0.193 & 1.043 \\
  0.252 & 0.866 & 2.550 & 0.473 & 0.647 & 0.232 & 0.878 & 0.193 & 1.044 \\
  0.203 & 0.877 & 2.561 & 0.478 & 0.646 & 0.228 & 0.874 & 0.190 & 1.036 \\
  0.157 & 0.887 & 2.580 & 0.483 & 0.649 & 0.231 & 0.880 & 0.194 & 1.046 \\
  0.113 & 0.894 & 2.575 & 0.486 & 0.644 & 0.224 & 0.868 & 0.187 & 1.024 \\
  0.073 & 0.899 & 2.581 & 0.488 & 0.646 & 0.224 & 0.870 & 0.189 & 1.029 \\
  0.034 & 0.912 & 2.586 & 0.494 & 0.642 & 0.217 & 0.859 & 0.181 & 1.008 \\
  0.000 & 0.920 & 2.604 & 0.498 & 0.645 & 0.219 & 0.864 & 0.183 & 1.015 \\
  \hline
\end{array}
$$
\end{center}
\end{table*}
 
\begin{table*}
\begin{center}
\caption{K corrections for the accretion model.}
$$
\begin{array}{rrrrrrrrrrr} \hline
  z & K_{U} & K_{B} & K_{V} & K_{R} & K_{I} & K_{J} & K_{H} & K_{K} &
  K_{L} & K_{M} \\ \hline
 4.035 &10.511 & 4.784 & 2.317 & 1.765 & 1.477 & 1.376 & 1.362 & 0.545 &  0.025 & -1.321 \\
 3.150 & 4.402 & 2.390 & 1.752 & 1.696 & 1.734 & 1.566 & 0.985 & 0.574 & -0.047 & -1.422 \\
 2.569 & 1.826 & 1.860 & 1.670 & 1.735 & 1.785 & 1.579 & 0.900 & 0.541 & -0.085 & -1.514 \\
 2.153 & 1.385 & 1.600 & 1.644 & 1.723 & 1.792 & 1.152 & 0.902 & 0.491 & -0.083 & -1.246 \\
 1.837 & 1.191 & 1.614 & 1.712 & 1.803 & 1.832 & 1.001 & 0.896 & 0.494 & -0.091 & -1.186 \\
 1.586 & 1.110 & 1.647 & 1.791 & 1.868 & 1.849 & 1.052 & 0.950 & 0.539 & -0.096 & -1.171 \\
 1.382 & 1.173 & 1.724 & 1.933 & 1.952 & 1.918 & 1.145 & 1.052 & 0.616 & -0.084 & -1.128 \\
 1.211 & 1.151 & 1.756 & 1.960 & 1.945 & 1.841 & 1.176 & 1.114 & 0.687 & -0.078 & -1.002 \\
 1.065 & 1.199 & 1.875 & 2.019 & 1.964 & 1.615 & 1.176 & 1.117 & 0.702 & -0.121 & -0.906 \\
 0.939 & 1.190 & 1.878 & 1.976 & 1.875 & 1.385 & 1.136 & 1.061 & 0.674 & -0.140 & -0.789 \\
 0.828 & 1.217 & 1.883 & 1.932 & 1.745 & 1.246 & 1.066 & 0.975 & 0.602 & -0.167 & -0.789 \\
 0.730 & 1.240 & 1.857 & 1.858 & 1.575 & 1.168 & 0.992 & 0.888 & 0.512 & -0.178 & -0.888 \\
 0.642 & 1.218 & 1.773 & 1.750 & 1.373 & 1.074 & 0.899 & 0.802 & 0.421 & -0.184 & -0.853 \\
 0.562 & 1.180 & 1.671 & 1.593 & 1.160 & 0.984 & 0.802 & 0.724 & 0.330 & -0.166 & -0.764 \\
 0.489 & 1.098 & 1.546 & 1.379 & 0.976 & 0.883 & 0.704 & 0.654 & 0.243 & -0.110 & -0.736 \\
 0.423 & 1.028 & 1.451 & 1.158 & 0.851 & 0.781 & 0.608 & 0.585 & 0.150 & -0.056 & -0.772 \\
 0.362 & 0.952 & 1.336 & 0.926 & 0.747 & 0.681 & 0.515 & 0.520 & 0.075 & -0.045 & -0.771 \\
 0.305 & 0.849 & 1.198 & 0.739 & 0.641 & 0.590 & 0.427 & 0.447 & 0.016 & -0.063 & -0.707 \\
 0.252 & 0.754 & 1.045 & 0.601 & 0.543 & 0.505 & 0.347 & 0.382 & -0.016 & -0.073 & -0.605 \\
 0.203 & 0.653 & 0.871 & 0.489 & 0.455 & 0.423 & 0.280 & 0.318 & -0.032 & -0.057 & -0.526 \\
 0.157 & 0.552 & 0.686 & 0.380 & 0.362 & 0.342 & 0.220 & 0.253 & -0.032 & -0.026 & -0.462 \\
 0.113 & 0.449 & 0.495 & 0.276 & 0.265 & 0.251 & 0.160 & 0.186 & -0.033 & -0.001 & -0.365 \\
 0.073 & 0.335 & 0.310 & 0.186 & 0.170 & 0.166 & 0.107 & 0.120 & -0.030 & 0.004 & -0.246 \\
 0.034 & 0.185 & 0.139 & 0.094 & 0.082 & 0.081 & 0.051 & 0.051 & -0.021 & -0.005 & -0.110 \\
 0.000 & 0.000 & 0.000 & 0.000 & 0.000 & 0.000 & 0.000 & 0.000 & 0.000 & 0.000 & -0.000 \\
  \hline
\end{array}
$$
\end{center}
\end{table*}
 
\begin{table*}
\begin{center}
\caption{Apparent integrated magnitudes and colours of the accretion
model.}
$$
\begin{array}{rrrrrrrr} \hline
 z &  m-M & m_{Bol} &  m_{U} &  m_{V} &  m_{K} &  U-B &  B-V \\ \hline
 4.035 &  47.795 &  31.921 &  39.855 &  32.253 &  29.668 &  5.177 &  2.425 \\
 3.150 &  47.153 &  30.267 &  32.478 &  30.281 &  28.211 &  1.543 &  0.654 \\
 2.569 &  46.619 &  28.489 &  28.484 &  28.708 &  26.559 & -0.473 &  0.249 \\
 2.153 &  46.153 &  26.778 &  26.639 &  27.229 &  24.958 & -0.633 &  0.043 \\
 1.837 &  45.731 &  25.398 &  25.276 &  26.058 &  23.591 & -0.810 &  0.028 \\
 1.586 &  45.338 &  24.259 &  24.246 &  25.128 &  22.425 & -0.902 &  0.020 \\
 1.382 &  44.969 &  23.453 &  23.677 &  24.547 &  21.494 & -0.877 &  0.007 \\
 1.211 &  44.615 &  22.791 &  23.195 &  24.047 &  20.765 & -0.904 &  0.052 \\
 1.065 &  44.269 &  22.253 &  22.879 &  23.646 &  20.124 & -0.928 &  0.161 \\
 0.939 &  43.931 &  21.721 &  22.520 &  23.193 &  19.557 & -0.914 &  0.241 \\
 0.828 &  43.595 &  21.216 &  22.200 &  22.746 &  19.012 & -0.866 &  0.320 \\
 0.730 &  43.258 &  20.745 &  21.895 &  22.299 &  18.497 & -0.797 &  0.393 \\
 0.642 &  42.918 &  20.272 &  21.543 &  21.830 &  18.013 & -0.723 &  0.436 \\
 0.562 &  42.567 &  19.783 &  21.131 &  21.281 &  17.523 & -0.652 &  0.502 \\
 0.489 &  42.203 &  19.284 &  20.667 &  20.679 &  17.037 & -0.610 &  0.598 \\
 0.423 &  41.828 &  18.789 &  20.211 &  20.048 &  16.529 & -0.574 &  0.737 \\
 0.362 &  41.430 &  18.305 &  19.757 &  19.418 &  16.036 & -0.525 &  0.864 \\
 0.305 &  40.997 &  17.756 &  19.210 &  18.768 &  15.513 & -0.482 &  0.924 \\
 0.252 &  40.522 &  17.209 &  18.666 &  18.159 &  14.992 & -0.414 &  0.921 \\
 0.203 &  39.993 &  16.592 &  18.052 &  17.514 &  14.432 & -0.332 &  0.870 \\
 0.157 &  39.375 &  15.914 &  17.375 &  16.810 &  13.818 & -0.240 &  0.805 \\
 0.113 &  38.600 &  15.053 &  16.502 &  15.915 &  13.031 & -0.141 &  0.728 \\
 0.073 &  37.592 &  13.985 &  15.405 &  14.833 &  12.036 & -0.067 &  0.639 \\
 0.034 &  35.872 &  12.190 &  13.562 &  13.023 &  10.322 & -0.035 &  0.574 \\
 0.000 & 25.000^{1}&   1.291 &   2.563 &   2.089 &  -0.515 & -0.067 &  0.541 \\
\hline
\end{array}
$$
\end{center}
{\tiny $^{1}$To calculate apparent magnitudes for the zero redshift
model, a distance modulus of 25 was chosen, corresponding to a
distance of 1Mpc.}
\end{table*}
 
\begin{table*}
\begin{center}
\caption{Apparent integrated magnitudes and colours of the accretion
model (part 2).}
$$
\begin{array}{rrrrrrrrr} \hline
  z & V-I & V-K & R-I & J-H & H-K & J-K & K-L & K-M \\ \hline
 4.035 & 1.128 & 2.585 & 0.461 & 0.326 & 0.928 & 1.254 & 0.640 & 2.411 \\
 3.150 & 0.353 & 2.070 & 0.149 & 0.896 & 0.517 & 1.412 & 0.772 & 2.578 \\
 2.569 & 0.261 & 2.149 & 0.157 & 1.022 & 0.465 & 1.487 & 0.771 & 2.640 \\
 2.153 & 0.252 & 2.271 & 0.151 & 0.617 & 0.522 & 1.139 & 0.707 & 2.320 \\
 1.837 & 0.315 & 2.467 & 0.211 & 0.498 & 0.519 & 1.016 & 0.714 & 2.278 \\
 1.586 & 0.425 & 2.703 & 0.289 & 0.540 & 0.548 & 1.089 & 0.771 & 2.370 \\
 1.382 & 0.568 & 3.053 & 0.346 & 0.594 & 0.615 & 1.209 & 0.859 & 2.540 \\
 1.211 & 0.730 & 3.282 & 0.448 & 0.630 & 0.659 & 1.289 & 0.964 & 2.665 \\
 1.065 & 1.073 & 3.522 & 0.725 & 0.668 & 0.675 & 1.342 & 1.043 & 2.683 \\
 0.939 & 1.302 & 3.636 & 0.887 & 0.709 & 0.662 & 1.371 & 1.044 & 2.594 \\
 0.828 & 1.429 & 3.734 & 0.912 & 0.736 & 0.648 & 1.384 & 0.998 & 2.530 \\
 0.730 & 1.464 & 3.802 & 0.835 & 0.753 & 0.644 & 1.397 & 0.913 & 2.528 \\
 0.642 & 1.473 & 3.817 & 0.738 & 0.749 & 0.642 & 1.391 & 0.822 & 2.389 \\
 0.562 & 1.418 & 3.758 & 0.621 & 0.728 & 0.647 & 1.375 & 0.707 & 2.189 \\
 0.489 & 1.314 & 3.642 & 0.543 & 0.700 & 0.660 & 1.360 & 0.561 & 2.065 \\
 0.423 & 1.209 & 3.519 & 0.526 & 0.669 & 0.674 & 1.343 & 0.405 & 1.982 \\
 0.362 & 1.088 & 3.382 & 0.528 & 0.644 & 0.685 & 1.329 & 0.320 & 1.911 \\
 0.305 & 1.003 & 3.255 & 0.518 & 0.625 & 0.663 & 1.288 & 0.272 & 1.766 \\
 0.252 & 0.962 & 3.167 & 0.511 & 0.612 & 0.630 & 1.241 & 0.250 & 1.633 \\
 0.203 & 0.943 & 3.082 & 0.510 & 0.608 & 0.578 & 1.186 & 0.215 & 1.530 \\
 0.157 & 0.925 & 2.992 & 0.503 & 0.616 & 0.516 & 1.132 & 0.188 & 1.476 \\
 0.113 & 0.919 & 2.884 & 0.500 & 0.618 & 0.443 & 1.061 & 0.155 & 1.356 \\
 0.073 & 0.919 & 2.797 & 0.492 & 0.633 & 0.374 & 1.007 & 0.155 & 1.245 \\
 0.034 & 0.925 & 2.701 & 0.495 & 0.642 & 0.289 & 0.931 & 0.165 & 1.097 \\
 0.000 & 0.920 & 2.604 & 0.498 & 0.645 & 0.219 & 0.864 & 0.183 & 1.015 \\
  \hline
\end{array}
$$
\end{center}
\end{table*}
 
\begin{table*}
\begin{center}
\caption{Intrinsic integrated magnitudes and colours of the collapse
model.}
$$
\begin{array}{rrrrrrrr} \hline
  z & t [Gyr] & M_{Bol} &  M_{U} &  M_{V} &  M_{K} & (U-B)_{0} &
  (B-V)_{0}\\ \hline
 4.035 &  1.5 & -20.964 & -19.421 & -18.541 & -18.950 & -0.734 & -0.146 \\
 3.150 &  2.0 & -24.454 & -23.270 & -22.728 & -23.649 & -0.544 &  0.002 \\
 2.569 &  2.5 & -24.683 & -23.743 & -23.541 & -25.065 & -0.370 &  0.168 \\
 2.153 &  3.0 & -24.674 & -23.715 & -23.675 & -25.722 & -0.299 &  0.259 \\
 1.837 &  3.5 & -24.441 & -23.445 & -23.592 & -25.958 & -0.198 &  0.345 \\
 1.586 &  4.0 & -24.292 & -23.217 & -23.522 & -26.047 & -0.120 &  0.425 \\
 1.382 &  4.5 & -24.181 & -23.022 & -23.439 & -26.068 & -0.072 &  0.489 \\
 1.211 &  5.0 & -24.087 & -22.852 & -23.331 & -26.016 & -0.055 &  0.534 \\
 1.065 &  5.5 & -23.976 & -22.683 & -23.241 & -25.958 & -0.019 &  0.577 \\
 0.939 &  6.0 & -23.883 & -22.542 & -23.155 & -25.893 &  0.004 &  0.609 \\
 0.828 &  6.5 & -23.787 & -22.403 & -23.067 & -25.828 &  0.027 &  0.637 \\
 0.730 &  7.0 & -23.713 & -22.280 & -22.990 & -25.765 &  0.048 &  0.662 \\
 0.642 &  7.5 & -23.635 & -22.167 & -22.912 & -25.701 &  0.064 &  0.681 \\
 0.562 &  8.0 & -23.565 & -22.072 & -22.853 & -25.643 &  0.083 &  0.698 \\
 0.489 &  8.5 & -23.494 & -21.970 & -22.779 & -25.579 &  0.097 &  0.712 \\
 0.423 &  9.0 & -23.440 & -21.900 & -22.729 & -25.527 &  0.107 &  0.722 \\
 0.362 &  9.5 & -23.381 & -21.824 & -22.672 & -25.471 &  0.117 &  0.731 \\
 0.305 & 10.0 & -23.335 & -21.766 & -22.622 & -25.416 &  0.121 &  0.735 \\
 0.252 & 10.5 & -23.271 & -21.682 & -22.563 & -25.366 &  0.135 &  0.746 \\
 0.203 & 11.0 & -23.231 & -21.629 & -22.522 & -25.328 &  0.141 &  0.752 \\
 0.157 & 11.5 & -23.196 & -21.589 & -22.485 & -25.288 &  0.143 &  0.753 \\
 0.113 & 12.0 & -23.157 & -21.536 & -22.443 & -25.253 &  0.149 &  0.758 \\
 0.073 & 12.5 & -23.126 & -21.492 & -22.412 & -25.226 &  0.157 &  0.763 \\
 0.034 & 13.0 & -23.088 & -21.445 & -22.374 & -25.196 &  0.163 &  0.766 \\
 0.000 & 13.5 & -23.067 & -21.416 & -22.350 & -25.170 &  0.166 &  0.768 \\
  \hline
\end{array}
$$
\end{center}
\end{table*}
 
\begin{table*}
\begin{center}
\caption{Intrinsic integrated magnitudes and colours of the collapse
model (part 2).}
$$
\begin{array}{rrrrrrrrr} \hline
  z & (V-I)_{0} & (V-K)_{0} & (R-I)_{0} & (J-H)_{0} & (H-K)_{0} &
  (J-K)_{0} & (K-L)_{0} & (K-M)_{0} \\ \hline
 4.035 & 0.147 & 0.409 & 0.094 & 0.202 & 0.051 & 0.254 & 0.054 & 0.307 \\
 3.150 & 0.315 & 0.921 & 0.181 & 0.332 & 0.094 & 0.427 & 0.095 & 0.478 \\
 2.569 & 0.502 & 1.524 & 0.283 & 0.446 & 0.144 & 0.590 & 0.140 & 0.668 \\
 2.153 & 0.629 & 2.047 & 0.356 & 0.565 & 0.227 & 0.792 & 0.195 & 0.955 \\
 1.837 & 0.717 & 2.366 & 0.401 & 0.638 & 0.283 & 0.921 & 0.235 & 1.152 \\
 1.586 & 0.790 & 2.525 & 0.435 & 0.664 & 0.292 & 0.956 & 0.242 & 1.198 \\
 1.382 & 0.856 & 2.629 & 0.467 & 0.672 & 0.280 & 0.952 & 0.229 & 1.173 \\
 1.211 & 0.903 & 2.685 & 0.489 & 0.672 & 0.264 & 0.935 & 0.215 & 1.132 \\
 1.065 & 0.941 & 2.717 & 0.507 & 0.667 & 0.248 & 0.915 & 0.203 & 1.094 \\
 0.939 & 0.969 & 2.738 & 0.520 & 0.664 & 0.237 & 0.901 & 0.195 & 1.069 \\
 0.828 & 0.993 & 2.761 & 0.530 & 0.664 & 0.230 & 0.895 & 0.191 & 1.057 \\
 0.730 & 1.012 & 2.775 & 0.538 & 0.663 & 0.224 & 0.887 & 0.187 & 1.042 \\
 0.642 & 1.026 & 2.789 & 0.545 & 0.663 & 0.220 & 0.882 & 0.184 & 1.034 \\
 0.562 & 1.039 & 2.790 & 0.550 & 0.659 & 0.214 & 0.873 & 0.180 & 1.018 \\
 0.489 & 1.048 & 2.800 & 0.553 & 0.660 & 0.214 & 0.874 & 0.180 & 1.020 \\
 0.423 & 1.054 & 2.798 & 0.556 & 0.658 & 0.210 & 0.868 & 0.177 & 1.010 \\
 0.362 & 1.060 & 2.799 & 0.558 & 0.657 & 0.207 & 0.864 & 0.175 & 1.002 \\
 0.305 & 1.063 & 2.794 & 0.560 & 0.654 & 0.205 & 0.859 & 0.173 & 0.995 \\
 0.252 & 1.071 & 2.803 & 0.563 & 0.655 & 0.205 & 0.860 & 0.174 & 0.996 \\
 0.203 & 1.076 & 2.806 & 0.566 & 0.654 & 0.202 & 0.856 & 0.170 & 0.987 \\
 0.157 & 1.076 & 2.803 & 0.565 & 0.654 & 0.202 & 0.856 & 0.170 & 0.988 \\
 0.113 & 1.081 & 2.810 & 0.568 & 0.654 & 0.202 & 0.856 & 0.169 & 0.987 \\
 0.073 & 1.085 & 2.814 & 0.569 & 0.654 & 0.200 & 0.854 & 0.167 & 0.983 \\
 0.034 & 1.089 & 2.822 & 0.571 & 0.656 & 0.202 & 0.858 & 0.168 & 0.988 \\
 0.000 & 1.090 & 2.820 & 0.572 & 0.655 & 0.201 & 0.856 & 0.167 & 0.984 \\
  \hline
\end{array}
$$
\end{center}
\end{table*}
 
\begin{table*}
\begin{center}
\caption{K corrections for the collapse model.}
$$
\begin{array}{rrrrrrrrrrr} \hline
  z & K_{U} & K_{B} & K_{V} & K_{R} & K_{I} & K_{J} & K_{H} & K_{K} &
  K_{L} & K_{M} \\ \hline
 4.035 & 9.757 & 4.025 & 1.562 & 1.001 & 0.696 & 0.610 & 0.623 & 0.045 & -0.356 & -1.335 \\
 3.150 & 4.157 & 2.134 & 1.451 & 1.412 & 1.464 & 1.403 & 0.994 & 0.598 & -0.056 & -1.332 \\
 2.569 & 2.284 & 2.227 & 2.059 & 2.155 & 2.324 & 2.122 & 1.441 & 1.008 & 0.175 & -1.541 \\
 2.153 & 1.974 & 2.156 & 2.303 & 2.487 & 2.612 & 2.013 & 1.748 & 1.286 & 0.383 & -1.080 \\
 1.837 & 1.907 & 2.423 & 2.669 & 2.857 & 2.826 & 1.933 & 1.801 & 1.345 & 0.432 & -0.984 \\
 1.586 & 1.949 & 2.641 & 2.975 & 3.062 & 2.903 & 1.898 & 1.743 & 1.236 & 0.318 & -1.088 \\
 1.382 & 2.026 & 2.724 & 3.116 & 3.069 & 2.859 & 1.823 & 1.648 & 1.087 & 0.156 & -1.213 \\
 1.211 & 1.921 & 2.683 & 3.041 & 2.912 & 2.658 & 1.695 & 1.512 & 0.931 & -0.017 & -1.253 \\
 1.065 & 1.946 & 2.778 & 2.996 & 2.788 & 2.340 & 1.559 & 1.362 & 0.796 & -0.153 & -1.218 \\
 0.939 & 1.936 & 2.765 & 2.844 & 2.582 & 2.038 & 1.424 & 1.211 & 0.681 & -0.207 & -1.097 \\
 0.828 & 1.963 & 2.734 & 2.686 & 2.369 & 1.790 & 1.295 & 1.074 & 0.580 & -0.226 & -1.060 \\
 0.730 & 1.933 & 2.625 & 2.496 & 2.131 & 1.602 & 1.157 & 0.945 & 0.478 & -0.230 & -1.095 \\
 0.642 & 1.920 & 2.483 & 2.315 & 1.877 & 1.430 & 1.027 & 0.839 & 0.390 & -0.227 & -1.026 \\
 0.562 & 1.901 & 2.322 & 2.116 & 1.628 & 1.289 & 0.897 & 0.745 & 0.297 & -0.197 & -0.937 \\
 0.489 & 1.801 & 2.129 & 1.880 & 1.397 & 1.143 & 0.775 & 0.669 & 0.211 & -0.127 & -0.885 \\
 0.423 & 1.663 & 1.934 & 1.620 & 1.188 & 0.988 & 0.658 & 0.597 & 0.120 & -0.063 & -0.874 \\
 0.362 & 1.523 & 1.730 & 1.351 & 1.019 & 0.845 & 0.550 & 0.528 & 0.042 & -0.037 & -0.842 \\
 0.305 & 1.300 & 1.517 & 1.091 & 0.846 & 0.717 & 0.449 & 0.456 & -0.012 & -0.044 & -0.765 \\
 0.252 & 1.123 & 1.317 & 0.874 & 0.705 & 0.608 & 0.361 & 0.394 & -0.036 & -0.047 & -0.658 \\
 0.203 & 0.921 & 1.094 & 0.688 & 0.577 & 0.505 & 0.288 & 0.331 & -0.046 & -0.045 & -0.569 \\
 0.157 & 0.725 & 0.865 & 0.512 & 0.454 & 0.405 & 0.222 & 0.264 & -0.042 & -0.032 & -0.475 \\
 0.113 & 0.548 & 0.628 & 0.361 & 0.326 & 0.294 & 0.162 & 0.194 & -0.038 & -0.019 & -0.357 \\
 0.073 & 0.378 & 0.405 & 0.240 & 0.207 & 0.198 & 0.106 & 0.123 & -0.035 & -0.013 & -0.234 \\
 0.034 & 0.192 & 0.189 & 0.119 & 0.097 & 0.094 & 0.052 & 0.053 & -0.023 & -0.009 & -0.102 \\
 0.000 & 0.000 & 0.000 & 0.000 & 0.000 & 0.000 & 0.000 & 0.000 & 0.000 & 0.000 & 0.000 \\
  \hline
\end{array}
$$
\end{center}
\end{table*}
 
\begin{table*}
\begin{center}
\caption{Apparent integrated magnitudes and colours of the collapse
model.}
$$
\begin{array}{rrrrrrrr} \hline
  z &  m-M & m_{Bol} &  m_{U} &  m_{V} &  m_{K} &  U-B &  B-V \\ \hline
 4.035 &  47.795 &  30.731 &  38.131 &  30.816 &  28.890 &  4.998 &  2.317 \\
 3.150 &  47.153 &  25.972 &  28.040 &  25.876 &  24.102 &  1.479 &  0.685 \\
 2.569 &  46.619 &  24.766 &  25.160 &  25.137 &  22.562 & -0.313 &  0.336 \\
 2.153 &  46.153 &  24.008 &  24.412 &  24.781 &  21.717 & -0.481 &  0.112 \\
 1.837 &  45.731 &  23.563 &  24.193 &  24.808 &  21.118 & -0.714 &  0.099 \\
 1.586 &  45.338 &  23.113 &  24.070 &  24.791 &  20.527 & -0.812 &  0.091 \\
 1.382 &  44.969 &  22.669 &  23.973 &  24.646 &  19.988 & -0.770 &  0.097 \\
 1.211 &  44.615 &  22.249 &  23.684 &  24.325 &  19.530 & -0.817 &  0.176 \\
 1.065 &  44.269 &  21.867 &  23.532 &  24.024 &  19.107 & -0.851 &  0.359 \\
 0.939 &  43.931 &  21.484 &  23.325 &  23.620 &  18.719 & -0.825 &  0.530 \\
 0.828 &  43.595 &  21.115 &  23.155 &  23.214 &  18.347 & -0.744 &  0.685 \\
 0.730 &  43.258 &  20.733 &  22.911 &  22.764 &  17.971 & -0.644 &  0.791 \\
 0.642 &  42.918 &  20.357 &  22.671 &  22.321 &  17.607 & -0.499 &  0.849 \\
 0.562 &  42.567 &  19.967 &  22.396 &  21.830 &  17.221 & -0.338 &  0.904 \\
 0.489 &  42.203 &  19.571 &  22.034 &  21.304 &  16.835 & -0.231 &  0.961 \\
 0.423 &  41.828 &  19.153 &  21.591 &  20.719 &  16.421 & -0.164 &  1.036 \\
 0.362 &  41.430 &  18.718 &  21.129 &  20.109 &  16.001 & -0.090 &  1.110 \\
 0.305 &  40.997 &  18.239 &  20.531 &  19.466 &  15.569 & -0.096 &  1.161 \\
 0.252 &  40.522 &  17.738 &  19.963 &  18.833 &  15.120 & -0.059 &  1.189 \\
 0.203 &  39.993 &  17.162 &  19.285 &  18.159 &  14.619 & -0.032 &  1.158 \\
 0.157 &  39.375 &  16.495 &  18.511 &  17.402 &  14.045 &  0.003 &  1.106 \\
 0.113 &  38.600 &  15.674 &  17.612 &  16.518 &  13.309 &  0.069 &  1.025 \\
 0.073 &  37.592 &  14.618 &  16.478 &  15.420 &  12.331 &  0.130 &  0.928 \\
 0.034 &  35.872 &  12.857 &  14.619 &  13.617 &  10.653 &  0.166 &  0.836 \\
 0.000 & 25.000^{1}&   1.933 &   3.584 &   2.650 &  -0.170 &  0.166 &  0.768 \\
\hline
\end{array}
$$
\end{center}
{\tiny $^{1}$To calculate apparent magnitudes for the zero redshift
model, a distance modulus of 25 was chosen, corresponding to a
distance of 1Mpc.}
\end{table*}
 
\begin{table*}
\begin{center}
\caption{Apparent integrated magnitudes and colours of the collapse
model (part 2).}
\label{liste10}
$$
\begin{array}{rrrrrrrrr} \hline
  z & V-I & V-K & R-I & J-H & H-K & J-K & K-L & K-M \\ \hline
 4.035 & 1.013 & 1.926 & 0.399 & 0.189 & 0.629 & 0.819 & 0.455 & 1.687 \\
 3.150 & 0.302 & 1.774 & 0.129 & 0.741 & 0.490 & 1.232 & 0.749 & 2.408 \\
 2.569 & 0.237 & 2.575 & 0.114 & 1.127 & 0.577 & 1.704 & 0.973 & 3.217 \\
 2.153 & 0.320 & 3.064 & 0.231 & 0.830 & 0.689 & 1.519 & 1.098 & 3.321 \\
 1.837 & 0.560 & 3.690 & 0.432 & 0.770 & 0.739 & 1.509 & 1.148 & 3.481 \\
 1.586 & 0.862 & 4.264 & 0.594 & 0.819 & 0.799 & 1.618 & 1.160 & 3.522 \\
 1.382 & 1.113 & 4.658 & 0.677 & 0.847 & 0.841 & 1.688 & 1.160 & 3.473 \\
 1.211 & 1.286 & 4.795 & 0.743 & 0.855 & 0.845 & 1.700 & 1.163 & 3.316 \\
 1.065 & 1.597 & 4.917 & 0.955 & 0.864 & 0.814 & 1.678 & 1.152 & 3.108 \\
 0.939 & 1.775 & 4.901 & 1.064 & 0.877 & 0.767 & 1.644 & 1.083 & 2.847 \\
 0.828 & 1.889 & 4.867 & 1.109 & 0.885 & 0.724 & 1.609 & 0.997 & 2.697 \\
 0.730 & 1.906 & 4.793 & 1.067 & 0.875 & 0.691 & 1.567 & 0.895 & 2.615 \\
 0.642 & 1.911 & 4.714 & 0.992 & 0.851 & 0.669 & 1.520 & 0.801 & 2.450 \\
 0.562 & 1.866 & 4.609 & 0.889 & 0.811 & 0.662 & 1.473 & 0.674 & 2.252 \\
 0.489 & 1.785 & 4.469 & 0.807 & 0.766 & 0.672 & 1.438 & 0.518 & 2.116 \\
 0.423 & 1.686 & 4.298 & 0.756 & 0.719 & 0.687 & 1.406 & 0.360 & 2.004 \\
 0.362 & 1.566 & 4.108 & 0.732 & 0.679 & 0.693 & 1.372 & 0.254 & 1.886 \\
 0.305 & 1.437 & 3.897 & 0.689 & 0.647 & 0.673 & 1.319 & 0.205 & 1.748 \\
 0.252 & 1.337 & 3.713 & 0.660 & 0.622 & 0.635 & 1.257 & 0.185 & 1.618 \\
 0.203 & 1.259 & 3.540 & 0.638 & 0.611 & 0.579 & 1.189 & 0.169 & 1.510 \\
 0.157 & 1.183 & 3.357 & 0.614 & 0.612 & 0.508 & 1.120 & 0.160 & 1.421 \\
 0.113 & 1.148 & 3.209 & 0.600 & 0.622 & 0.434 & 1.055 & 0.150 & 1.306 \\
 0.073 & 1.127 & 3.089 & 0.578 & 0.637 & 0.358 & 0.995 & 0.145 & 1.182 \\
 0.034 & 1.114 & 2.964 & 0.574 & 0.655 & 0.278 & 0.932 & 0.154 & 1.067 \\
 0.000 & 1.090 & 2.820 & 0.572 & 0.655 & 0.201 & 0.856 & 0.167 & 0.984 \\
  \hline
\end{array}
$$
\end{center}
\end{table*}

\section{Bulge colours}

\begin{table*}
\begin{center}
\caption{Bulge colours of the accretion model, face-on.}
\label{liste11}
$$
\begin{array}{rrrrrrrrrrr} \hline
  z & U-B & B-V & V-I & V-K & R-I & J-H & H-K & J-K & K-L & K-M \\ \hline
 4.035 & 5.411 & 2.515 & 1.182 & 2.809 & 0.490 & 0.374 & 1.006 & 1.380 & 0.659 & 2.582 \\
 3.150 & 2.013 & 0.819 & 0.569 & 2.889 & 0.240 & 1.166 & 0.605 & 1.771 & 0.837 & 3.038 \\
 2.569 & -0.176 & 0.571 & 0.571 & 3.508 & 0.328 & 1.333 & 0.634 & 1.966 & 0.883 & 3.308 \\
 2.153 & -0.354 & 0.195 & 0.449 & 3.194 & 0.275 & 0.787 & 0.649 & 1.436 & 0.881 & 2.897 \\
 1.837 & -0.691 & 0.081 & 0.512 & 3.053 & 0.397 & 0.607 & 0.615 & 1.222 & 0.915 & 2.795 \\
 1.586 & -0.808 & 0.033 & 0.771 & 3.585 & 0.504 & 0.674 & 0.722 & 1.395 & 1.021 & 3.056 \\
 1.382 & -0.869 & -0.070 & 0.889 & 4.071 & 0.470 & 0.798 & 0.951 & 1.749 & 1.266 & 3.697 \\
 1.211 & -0.867 & 0.249 & 1.340 & 4.764 & 0.691 & 0.819 & 0.964 & 1.783 & 1.315 & 3.673 \\
 1.065 & -0.863 & 0.302 & 1.369 & 4.348 & 0.863 & 0.783 & 0.853 & 1.637 & 1.220 & 3.208 \\
 0.939 & -0.764 & 0.570 & 1.881 & 4.777 & 1.136 & 0.820 & 0.761 & 1.581 & 1.095 & 2.821 \\
 0.828 & -0.732 & 0.703 & 1.932 & 4.714 & 1.143 & 0.856 & 0.742 & 1.598 & 1.045 & 2.753 \\
 0.730 & -0.843 & 0.352 & 1.416 & 3.926 & 0.848 & 0.832 & 0.746 & 1.577 & 1.065 & 2.891 \\
 0.642 & -0.759 & 0.522 & 1.636 & 4.319 & 0.846 & 0.889 & 0.757 & 1.645 & 0.973 & 2.817 \\
 0.562 & -0.641 & 0.606 & 1.644 & 4.474 & 0.760 & 0.913 & 0.772 & 1.684 & 0.847 & 2.679 \\
 0.489 & -0.550 & 0.688 & 1.569 & 4.481 & 0.682 & 0.911 & 0.789 & 1.700 & 0.670 & 2.589 \\
 0.423 & -0.578 & 0.737 & 1.458 & 4.289 & 0.648 & 0.844 & 0.779 & 1.623 & 0.457 & 2.388 \\
 0.362 & -0.438 & 0.929 & 1.392 & 4.265 & 0.659 & 0.825 & 0.805 & 1.630 & 0.349 & 2.326 \\
 0.305 & -0.493 & 0.949 & 1.202 & 3.933 & 0.599 & 0.764 & 0.769 & 1.533 & 0.278 & 2.106 \\
 0.252 & -0.461 & 0.961 & 1.107 & 3.857 & 0.577 & 0.759 & 0.763 & 1.521 & 0.279 & 2.041 \\
 0.203 & -0.391 & 0.938 & 1.071 & 3.757 & 0.574 & 0.742 & 0.710 & 1.452 & 0.256 & 1.925 \\
 0.157 & -0.385 & 0.805 & 0.963 & 3.515 & 0.532 & 0.732 & 0.632 & 1.364 & 0.234 & 1.805 \\
 0.113 & -0.289 & 0.731 & 0.980 & 3.540 & 0.544 & 0.767 & 0.585 & 1.352 & 0.235 & 1.768 \\
 0.073 & -0.062 & 0.744 & 1.065 & 3.556 & 0.569 & 0.776 & 0.495 & 1.271 & 0.218 & 1.609 \\
 0.034 & 0.071 & 0.699 & 1.086 & 3.443 & 0.579 & 0.784 & 0.398 & 1.182 & 0.229 & 1.451 \\
 0.000 & -0.049 & 0.586 & 1.034 & 3.199 & 0.566 & 0.766 & 0.298 & 1.064 & 0.238 & 1.304 \\
  \hline
\end{array}
$$
\end{center}
\end{table*}
 
\begin{table*}
\begin{center}
\caption{Bulge colours of the accretion model, edge-on.}
\label{liste12}
$$
\begin{array}{rrrrrrrrrrr} \hline
 z & U-B & B-V & V-I & V-K & R-I & J-H & H-K & J-K & K-L & K-M \\ \hline
 4.035 & 5.574 & 2.562 & 1.211 & 2.911 & 0.504 & 0.395 & 1.038 & 1.433 & 0.671 & 2.640 \\
 3.150 & 2.332 & 0.868 & 0.623 & 3.047 & 0.263 & 1.207 & 0.620 & 1.827 & 0.849 & 3.104 \\
 2.569 & -0.295 & 0.471 & 0.497 & 3.304 & 0.289 & 1.298 & 0.628 & 1.926 & 0.882 & 3.280 \\
 2.153 & -0.521 & 0.102 & 0.334 & 2.758 & 0.209 & 0.732 & 0.619 & 1.350 & 0.857 & 2.796 \\
 1.837 & -0.706 & 0.063 & 0.499 & 2.956 & 0.398 & 0.600 & 0.618 & 1.218 & 0.944 & 2.830 \\
 1.586 & -0.782 & -0.017 & 0.968 & 4.126 & 0.623 & 0.776 & 0.841 & 1.617 & 1.172 & 3.500 \\
 1.382 & -0.816 & -0.021 & 1.114 & 4.730 & 0.585 & 0.885 & 1.089 & 1.974 & 1.487 & 4.245 \\
 1.211 & -0.955 & 0.109 & 1.038 & 4.731 & 0.563 & 0.942 & 1.190 & 2.133 & 1.705 & 4.666 \\
 1.065 & -0.945 & 0.341 & 1.372 & 4.981 & 0.868 & 0.991 & 1.198 & 2.189 & 1.834 & 4.795 \\
 0.939 & -0.906 & 0.805 & 1.768 & 5.531 & 1.088 & 1.151 & 1.202 & 2.353 & 1.818 & 4.785 \\
 0.828 & -0.845 & 0.498 & 1.462 & 5.140 & 0.926 & 1.208 & 1.169 & 2.377 & 1.749 & 4.739 \\
 0.730 & -0.596 & 0.669 & 1.737 & 5.595 & 0.960 & 1.264 & 1.252 & 2.516 & 1.814 & 5.135 \\
 0.642 & -0.500 & 0.582 & 1.631 & 5.704 & 0.832 & 1.337 & 1.301 & 2.639 & 1.744 & 5.158 \\
 0.562 & -0.553 & 0.511 & 1.532 & 5.598 & 0.723 & 1.304 & 1.276 & 2.580 & 1.558 & 4.830 \\
 0.489 & -0.531 & 0.560 & 1.427 & 5.447 & 0.655 & 1.250 & 1.265 & 2.515 & 1.361 & 4.696 \\
 0.423 & -0.304 & 0.891 & 1.533 & 6.062 & 0.736 & 1.373 & 1.414 & 2.787 & 1.280 & 4.893 \\
 0.362 & -0.538 & 0.850 & 1.405 & 5.839 & 0.738 & 1.315 & 1.341 & 2.657 & 0.965 & 4.516 \\
 0.305 & -0.213 & 1.174 & 1.456 & 5.956 & 0.780 & 1.300 & 1.322 & 2.622 & 0.909 & 4.406 \\
 0.252 & -0.416 & 0.980 & 1.302 & 5.757 & 0.718 & 1.298 & 1.292 & 2.590 & 0.839 & 4.269 \\
 0.203 & -0.393 & 0.881 & 1.267 & 5.740 & 0.705 & 1.313 & 1.262 & 2.575 & 0.831 & 4.194 \\
 0.157 & -0.185 & 0.912 & 1.344 & 5.784 & 0.739 & 1.319 & 1.121 & 2.440 & 0.673 & 3.775 \\
 0.113 & -0.247 & 0.753 & 1.327 & 5.768 & 0.744 & 1.340 & 1.066 & 2.406 & 0.678 & 3.748 \\
 0.073 & 0.161 & 0.891 & 1.498 & 5.742 & 0.814 & 1.298 & 0.883 & 2.181 & 0.538 & 3.149 \\
 0.034 & 0.194 & 0.884 & 1.636 & 5.797 & 0.898 & 1.286 & 0.773 & 2.059 & 0.557 & 2.917 \\
 0.000 & 0.030 & 0.830 & 1.644 & 5.551 & 0.920 & 1.244 & 0.629 & 1.872 & 0.503 & 2.623 \\
  \hline
\end{array}
$$
\end{center}
\end{table*}

\end{document}